\documentclass[12pt,letterpaper]{article}
\usepackage[utf8]{inputenc}

\usepackage{graphicx,array}
\usepackage{url}
\usepackage{color}
\usepackage{latexsym}
\usepackage{amsthm}
\usepackage{amsmath}
\usepackage{amssymb}
\usepackage{amsfonts}
\usepackage[numbers,sort&compress]{natbib}
\usepackage{bm}
\usepackage{bbm}
\usepackage{slashed}
\usepackage{mathrsfs}
\usepackage{enumerate}
\usepackage{tikz}
\usepackage{siunitx}
\usepackage{mdframed}
\usepackage{setspace}  
\usepackage{esvect}
\usepackage{physics}
\usepackage{cancel}
\usepackage{subfigure}
\usepackage{enumitem}
\usepackage{booktabs}
\usepackage{tcolorbox}%

\numberwithin{equation}{section}

\usepackage{hyperref} 
\hypersetup{
    colorlinks=true,       
    linkcolor=red,          
    citecolor=blue,        
    filecolor=magenta,      
    urlcolor=blue           
}
\usepackage[all]{hypcap} 
\usepackage{multirow}
\usepackage{multicol}

\newcolumntype{M}[1]{>{\centering\arraybackslash}p{#1}}

\usepackage{natbib}
\setlist[description]{leftmargin=\parindent,labelindent=\parindent}

\usepackage[normalem]{ulem} 

\newcommand{\nc}{\newcommand}

\nc{\beq}{\begin{equation}}
\nc{\eeq}{\end{equation}}
\nc{\beqa}{\begin{eqnarray}}  
\nc{\eeqa}{\end{eqnarray}}  
\nc{\bit}{\begin{itemize}}  
\nc{\eit}{\end{itemize}}

\usepackage{floatrow}
\newfloatcommand{capbtabbox}{table}[][\FBwidth]

\usepackage{blindtext}

\allowdisplaybreaks
\title{
{\bf Heterogeneous Cosmological Phase Transitions:}\\
{\bf \Large Seeded by Domain Walls and Junctions}
\author{\large Yang Bai$^{\,a,b}$, Yifu Xu$^{c}$, and Yiming Yang$^{a}$}
\date{\small \it 
$^a$Department of Physics, University of Wisconsin-Madison, Madison, WI 53706, USA \\
$^b$HEP Division, Argonne National Laboratory, 
Argonne, IL 60439, USA \\
$^c$University of Oxford, Oxford OX1 9FB, United Kingdom
}}

\begin{document}

\maketitle

\begin{abstract}
Heterogeneous nucleation is central to many familiar first-order phase transitions such as the freezing of water and the solidification of metals, and it can also play a crucial role in cosmology. We examine nucleation seeded by preexisting domain walls and demonstrate its strong impact on the dynamics of cosmological phase transitions. The bubble solutions take the form of spherical caps, and the contact angle is fixed by the ratio of the domain-wall tension to the bubble-wall tension. A larger domain-wall tension, or equivalently a smaller contact angle, reduces the wall-seeded bubble volume and lowers the critical nucleation action. For theories with $\mathbb{Z}_{n\geq 3}$ symmetry, domain-wall junctions naturally appear and we find that they seed nucleation even more efficiently than the walls themselves. Using a two-scalar-field model as an illustration, we compute nucleation temperatures for both homogeneous and heterogeneous channels and show that junction-seeded nucleation occurs at a higher temperature and is the dominant mechanism that completes the first-order cosmological phase transition. \end{abstract}

\thispagestyle{empty}  
\newpage    
\setcounter{page}{1}  

\begingroup
\hypersetup{linkcolor=black,linktocpage}
\tableofcontents
\endgroup

\newpage


\section{Introduction}
Cosmological phase transitions play an important role in addressing open questions in the Standard Model (SM) of particle physics and cosmology, with first-order phase transitions (FOPTs) being a particularly prominent topic. A notable example is the electroweak FOPT, which may help explain the baryon–antibaryon asymmetry when supplemented by physics beyond the SM~\cite{Kuzmin:1985mm, Cohen:1993nk, Rubakov:1996vz}. Interest in this subject has grown further with the ongoing observations by the LIGO–Virgo gravitational-wave (GW) observatories~\cite{LIGOScientific:2016aoc} and the advent of next-generation GW detectors—ground-based facilities such as the Einstein Telescope~\cite{ET:2019dnz}; space-based missions including LISA~\cite{Caprini:2019egz}, TianQin~\cite{TianQin:2015yph}, and Taiji~\cite{Hu:2017mde}; and pulsar-timing-array experiments such as NANOGrav~\cite{NANOGrav:2023gor}. These instruments are expected to probe stochastic GW backgrounds in the near future~\cite{Athron:2023xlk}, potentially revealing signals originating from cosmological FOPTs.

Most existing studies of cosmological FOPTs in the literature focus on homogeneous nucleation, whose formulation is well established in quantum field theory. In this scenario, the FOPT proceeds via tunneling through a bounce solution, which corresponds to a saddle point of the Euclidean action. The foundational work on quantum tunneling was developed in~\cite{Coleman:1977py, Callan:1977pt}, while the theory of thermal tunneling was established in~\cite{Linde:1977mm, Linde:1980tt, Linde:1981zj}. The criterion for when a phase transition completes in the early universe was discussed in~\cite{Guth:1981uk}. In the homogeneous case, rotational symmetry is assumed in the equations of motion, leading to a spherically symmetric bounce corresponding to a spherical critical bubble.

However, heterogeneous phase transitions—where the transition is initiated at defects or impurities in the system—are in fact more common in everyday physics and condensed-matter systems. The earliest discussions of heterogeneous phase transitions can be traced back nearly a century to Volmer’s work~\cite{Volmer1929BERKU,Volmer_book} on solid–liquid–gas transitions, where he described how phase boundaries catalyze the nucleation of a more stable phase. This line of research was further developed by Turnbull and collaborators in studies of metal solidification~\cite{Turnbull_Fisher,Turnbull}. With the subsequent development of crystallography, defects received increasing attention in the context of crystallization theory and crystal growth~\cite{mullin2001crystallization}.

Impurities in the early universe can also induce FOPTs from a field-theoretic perspective, giving rise to heterogeneous cosmological phase transitions, which are the focus of this paper. Soliton solutions to the classical equations of motion—whether topological or non-topological—naturally serve as such defects in our context because they possess localized energy density. Early studies include Refs.~\cite{Steinhardt:1981ec, Hosotani:1982ii, Yajnik:1986tg} in general settings and works specifically addressing the confinement phase transition~\cite{Frei:1989es, Trappenberg:1992bt, Christiansen:1995ic, Holland:2000uj}. Other proposals suggest that compact objects~\cite{Hiscock:1987hn, Burda:2015isa, Oshita:2018ptr} or localized baryonic matter~\cite{Grinstein:2015jda} in the universe may also act as impurities. More recent papers have focused on topological-defect–induced phase transitions, including those seeded by domain walls~\cite{Blasi:2022woz, Agrawal:2023cgp} and cosmic strings~\cite{Blasi:2024mtc}.

In this study, we choose domain walls as an illustrative example for investigating heterogeneous cosmological phase transitions, owing to their appearance in many beyond-the-SM models. The presence of domain-wall networks inevitably leads to the formation of junctions, where multiple vacua meet at a single point—or, more precisely, along a short line segment. While junctions have been explored through dynamical simulations of their stability~\cite{PinaAvelino:2006ia,Avelino:2006xf,Avelino:2008ve,Battye:2011ff} and in various supersymmetric models~\cite{Abraham:1990nz,Gibbons:1999np,Carroll:1999wr}, their role in seeding cosmological phase transitions remains largely unexplored. In this paper, we systematically study both domain-wall– and junction–seeded phase transitions, with the latter going beyond recent analyses in Refs.~\cite{Blasi:2022woz, Agrawal:2023cgp}.

Unlike Refs.~\cite{Blasi:2022woz, Agrawal:2023cgp}, which introduced elliptic bubbles to model the critical configuration, we construct a simple physical model in which bubbles nucleated on defects are described by spherical caps with a contact angle connecting the bubble to a domain wall (or junction), following the approach of Refs.~\cite{Turnbull, Grinstein:2015jda}. This spherical‐cap geometry with a contact angle captures both the critical bounce action and the subsequent bubble expansion. We will show that the critical bubble actions for the different nucleation channels—homogeneous, wall-seeded, and junction-seeded—are proportional to their corresponding critical bubble volumes. Since the wall‐ and junction‐seeded configurations have smaller volumes than the homogeneous one, they dominate the dynamics of the FOPT.

As an illustrative example, we consider a model with one real scalar field $\phi$ and one complex scalar field $S$ charged under a $\mathbb{Z}_n$ symmetry, which exhibits a two-step thermal phase transition. At high temperatures, $S$ acquires a vacuum expectation value (VEV), spontaneously breaking the $\mathbb{Z}_n$ symmetry and generating domain walls and junctions. At a lower critical temperature $T_c$, a new global minimum develops along the $\phi$ direction while $S$ returns to a zero VEV. This triggers a FOPT in which the fields tunnel from the false vacuum ($\langle S\rangle\neq 0$, $\langle\phi\rangle = 0$) to the true vacuum ($\langle S\rangle = 0$, $\langle\phi\rangle \neq 0$), thereby restoring the $\mathbb{Z}_n$ symmetry.

The bounce profile for the second-step FOPT can be approximated by a spherical cap nucleated on the defect, characterized by a contact angle related—in the thin-wall limit—to the ratio of the domain-wall tension to the bubble-wall tension. We also introduce an improved approximate description of the profile using a $\tanh$-type interpolation, which better captures the field configuration beyond the thin-wall approximation. To validate our analytic approximations, we use the mountain-pass theorem (MPT) algorithm implemented in Refs.~\cite{Agrawal:2022hnf, Agrawal:2023cgp} to obtain numerical solutions for the wall-seeded bubble configuration.

Our paper is organized as follows. In Section~\ref{sec:general_nucleation}, we formulate the general framework for heterogeneous bubble nucleation on domain walls and junctions, modeling the critical bubble as a spherical cap characterized by a contact angle. In Section~\ref{sec:Heter-Cosmo}, we apply heterogeneous phase transitions to cosmology, detailing the conditions for nucleation and percolation. Section~\ref{sec:domain-wall} introduces $\mathbb{Z}_2$- and $\mathbb{Z}_3$‐symmetric scalar-field models and analyzes the domain-wall–seeded phase transition using several complementary methods. In Section~\ref{sec:junction}, we discuss junctions and the transitions they induce in models with $\mathbb{Z}_3$ symmetry. We conclude in Section~\ref{sec:conclusions}. Additional topics—including possible symmetry-breaking patterns of $\mathbb{Z}_2$, the stability of domain walls, and special constraints relevant to field theory—are summarized in Appendices~\ref{app:differentZ2}, \ref{app:stability}, and \ref{app:sigma_DW-sigma_B}, respectively.

\section{Heterogeneous nucleation} \label{sec:general_nucleation}

In this section, we first discuss some general properties of heterogeneous nucleation, in contrast to homogeneous nucleation. To make the presentation simpler—and also to align with the cosmological phase transition induced by domain walls—we consider a flat domain wall with bubble nucleation occurring on both sides of the wall (in contrast to boiling water, where bubbles typically start only on one side of a cooking pan).

Much of what is known about homogeneous and heterogeneous nucleation rates originates from studies of the solidification of liquid metal. For instance, by examining the thermal phase transition of $\mathcal{O}(10)$-micron liquid-metal droplets, which suppress impurity effects, one can establish the homogeneous nucleation rate using the action of a spherical critical bubble~\cite{Turnbull}
\beqa
\label{eq:S3-thin-wall-homo}
S_{3,c}^{\rm homo} \approx \frac{16\,\pi\,\sigma_{\rm B}^3}{3\,\Delta V^2} ~,
\eeqa
where $\sigma_{\rm B}$ is the bubble tension and $\Delta V$ is the effective potential difference between the inside and outside of the bubble. To derive this expression, one applies the ``thin-wall'' approximation, in which the wall thickness is much smaller than the bubble radius. In this limit, the action can be approximated as a function of the bubble radius $R$ as $S_3 \approx 4\pi R^2 \,\sigma_{\rm B} - \frac{4\pi}{3}\,R^3\,\Delta V$, and extremizing this expression yields the critical bubble size $R_{\rm c} = 2 \sigma_{\rm B}/\Delta V$ and the action in \eqref{eq:S3-thin-wall-homo}.  

\begin{figure}[th!]
    \centering
    \subfigure[]{\includegraphics[width=0.3\linewidth]{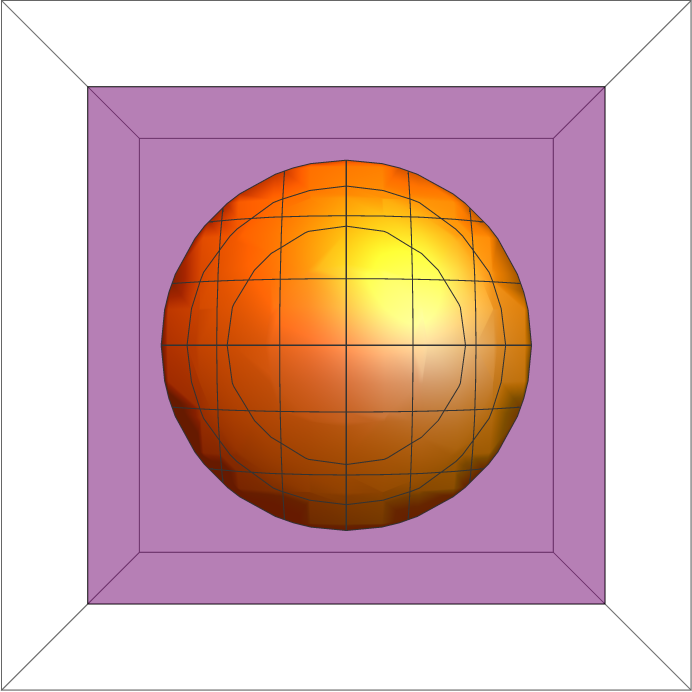}}
    \subfigure[]{\includegraphics[width=0.3\linewidth]{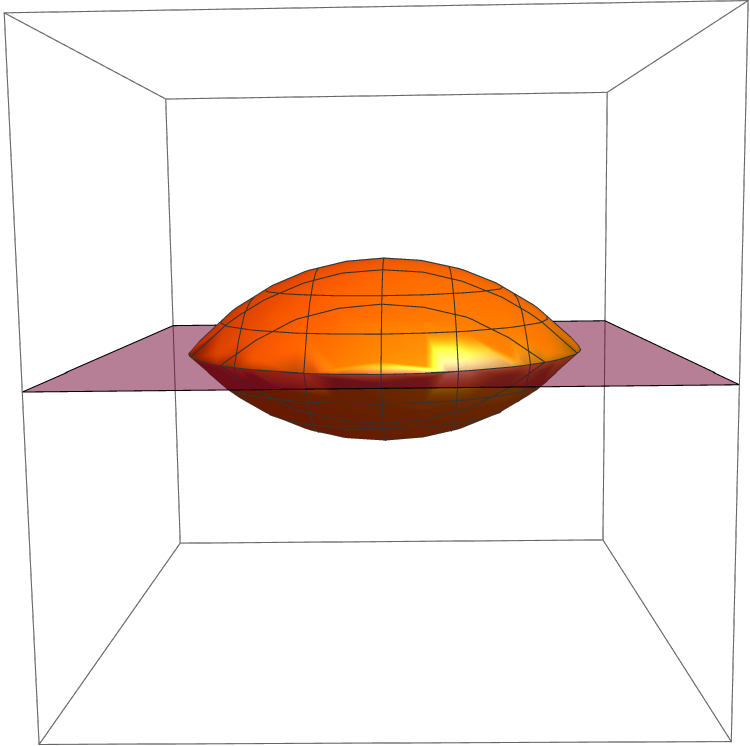}}
    \subfigure[]{\label{fig:DWbubbleModel_sphcap-c} \includegraphics[width=0.35\linewidth]{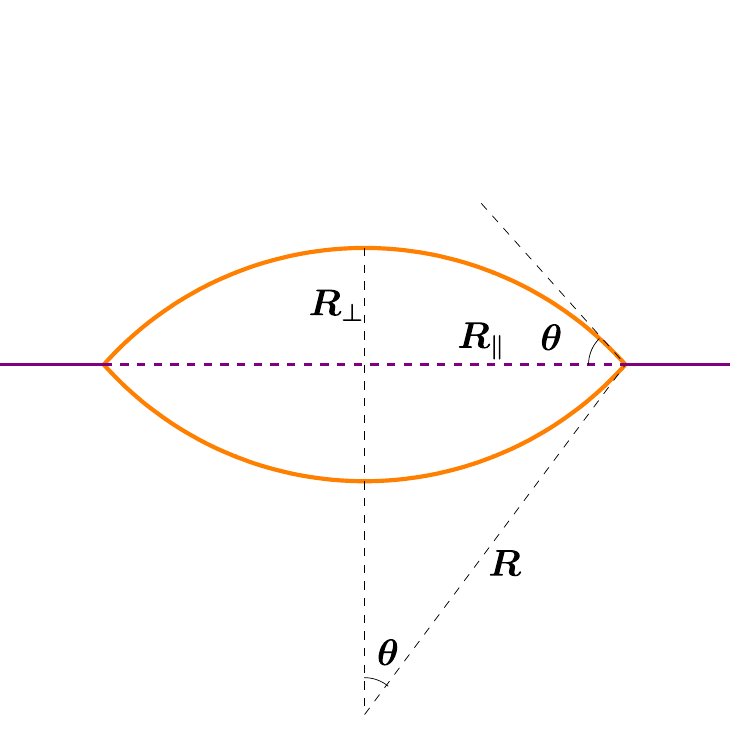}}
    
    \caption{Left panel: 3d visualization of a nucleated bubble attached to the domain wall (side view). Middle panel: top view. Right panel: notation for the top view, showing the orange spherical cap as the bubble wall and the purple straight line as the domain wall. The purple dashed line indicates the disappearance (``melting") of the domain wall inside the nucleated bubble. The contact angle is defined as $\theta$,  $R_\parallel = R \sin{\theta}$ and $R_\perp = R(1-\cos{\theta})$.
    }
    \label{fig:DWbubbleModel_sphcap}
\end{figure}

\subsection{Domain wall}

For heterogeneous nucleation, we first consider the simplest case of a flat domain wall with a bubble configuration illustrated in Fig.~\ref{fig:DWbubbleModel_sphcap}. Unlike the water-wetting scenario—which involves three interfaces among solid, liquid, and gas—there are only two interfaces in the domain-wall–seeded cosmological phase transition. This reduction arises because the domain wall effectively ``melts" or disappears inside the nucleated bubble.

Denoting the phase inside the bubble as the true vacuum (“T”) and the phase outside as the false vacuum (“F”), the two relevant interfaces are the “T–DW’’ interface (represented by the orange spherical cap with tension $\sigma_{\rm B}$) and the “F–DW’’ interface (represented by the purple straight segment with tension $\sigma_{\rm DW}$) in Fig.~\ref{fig:DWbubbleModel_sphcap}. The force balance at the contact line is then determined by the Young equation~\cite{young1805,gibbs1957collected}, which reads 
\beqa
\label{eq:young}
\sigma_{\text{DW}}=2\,\sigma_{\text{B}}\,\cos\theta ~,
\eeqa
with $\theta \le \pi/2$ the contact angle, defined in the right panel of Fig.~\ref{fig:DWbubbleModel_sphcap}. Note that in ordinary liquid wetting, one may have $\theta > \pi/2$ because the wall can persist inside the bubble. We do not consider more complicated configurations involving a higher-tension domain wall contained within the bubble.~\footnote{For certain field-theory models, the early, spontaneously broken discrete symmetry may be restored, while at the same time an additional discrete symmetry becomes spontaneously broken during the first-order phase transition. The original domain walls inside the bubble will then be replaced by a new domain wall with a different wall tension.}

The contribution to the action (more precisely, the action difference between the wall-seeded bubble configuration and the original domain-wall configuration) consists of three parts
\beqa
S_3^{\rm W} = 4\pi\,R^2\,(1-\cos{\theta})\,\sigma_{\rm B} - \frac{4\pi}{3}\,R^3\,(1-\cos{\theta})^2\,\left(1+\frac{\cos{\theta}}{2}\right)\,\Delta V - \pi R^2\,(1-\cos^2{\theta})\,\sigma_{\rm DW} ~,
\label{eqn:S3dw_thinwall}
\eeqa
where the first term is the contribution from the bubble wall, the second term comes from the bulk potential energy difference, and the third term arises from the disappearance of the domain wall inside the bubble. Treating both $R$ and $\theta$ as variation parameters, one can obtain the saddle point with $R_c = 2\sigma_{\rm B}/\Delta V$ (the same as the critical bubble radius of the homogeneous case) and $\cos{\theta} = \sigma_{\rm DW}/(2\sigma_{\rm B})$ [the same as the force balance condition in \eqref{eq:young}]. Note that the action is a minimum in $\theta$, but a maximum in $R$. The action of the critical configuration at the saddle point is 
\beqa
S^{\rm W}_{3,c} = \frac{16\,\pi\,\sigma_{\rm B}^3}{3\,\Delta V^2} \, (1-\cos{\theta})^2\,\left(1 + \frac{\cos{\theta}}{2}\right) ~. \label{eqn:S3dw_critical_thinwall}
\eeqa
The above formula was derived long ago in Ref.~\cite{Turnbull}, although here we have re-derived it using the extremalized-action method. In the limit $\sigma_{\rm DW} \ll \sigma_{\rm B}$ or $\cos{\theta} \rightarrow 0$, this action becomes identical to the homogeneous case. In the opposite limit, $\sigma_{\rm DW} \gg \sigma_{\rm B}$ or $\cos{\theta} \rightarrow 1$ (the ``complete wetting" limit), one has $S^{\rm W}_{3,c} \rightarrow 0$, meaning that it takes essentially no action for the domain wall to source the critical configuration. For comparable $\sigma_{\rm B}$ and $\sigma_{\rm DW}$, one finds $S^{\rm W}_{3,c} \le S^{\rm homo}_{3,c}$ for $\theta \le \pi/2$ or $\sigma_{\rm DW} \le 2\sigma_{\rm B}$. This indicates that the wall-seeded bubble, with the domain wall melted inside, has a lower action and therefore a larger nucleation rate. 

Also note the reduction factor is exactly the same as the ratio of the volume of the double-sphere-cap or the lenticular shape. This can be understood in two steps. First, the critical radius is always $R_c = 2 \sigma_{\rm B}/\Delta V$, which is guaranteed by the fact that $R_c$ is not only the extremal value of the action globally, but locally. For instance, consider a small sphere cap away from the contact line, the solid angle is the same for the tension and the vacuum energy part. One therefore obtains the same critical radius as the spherical bubble case. Second, extremalizing the action in terms of $R$, one has the critical action to be $\frac{1}{2}$ of the absolute value of the vacuum energy part [the second term in \eqref{eqn:S3dw_thinwall}]. So, the critical bubble action is exactly the product of the homogenous critical bubble action and the volume reduction of lenticular shape. We will encounter the similar conclusion later when we discuss junction-seeded bubbles.

Eq.~\eqref{eqn:S3dw_critical_thinwall} is valid for any heterogeneous phase transition occurring on a two-dimensional defect. It is worth emphasizing the assumptions used here: (i) the thickness of the wall is negligible compared to the bubble size, i.e., the temperature is close to the critical temperature $T_c$; and (ii) the spatial extent of the defect is much larger than the bubble so that the plane can be treated as locally flat. We will consider corrections to assumption (i) later in Section~\ref{sec:domain-wall}. 

\subsection{Domain wall junction}

For a general $\mathbb{Z}_n$ domain wall networks, one also anticipates existence of wall junction when two or more walls intersect at one line. For $\mathbb{Z}_2$ domain walls, the junction configuration is not stable, which will not be considered in this study~\footnote{The junctions of $\mathbb{Z}_2$ domain walls, if they survive until the phase-transition temperature, can also seed bubbles.}. For a general junction with multiple walls intersecting, the Young's equation is given by the vector tension-balance condition
\beqa
\sum_{i=1}^{n_{\rm w}} \vec{\sigma}_i = 0 ~,
\label{eq:Young's-equation}
\eeqa
where $n_{\rm w}$ is the number of walls. For $n_{\rm w} = 3$, one has the $Y$-type junction with the angles between two walls determined by the ratios of wall tensions. For the simple case with equal tensions, one has all three angles to be $2\pi/3$. For $n_{\rm w} = 4$, one has the $X$-type junctions. Again, for the simple case with equal tension, the intersecting angle is $\pi/2$. Depending on the tension relations as well as the number of distinguishable domains, the stability of junction configurations can be studied at least locally to minimize the tension-weight line length. 

Let us take the $\mathbb{Z}_4$ symmetry as an example. It has two distinct wall tensions, $\sigma_{\rm DW,1}$ and $\sigma_{\rm DW,2}$, and without loss of generality we assume $\sigma_{\rm DW,1} < \sigma_{\rm DW,2}$. The $X$-type junction admits two classes of domain-order configurations (ordered clockwise). Type I: 
(1,2,3,4) and 
(1,4,3,2), which are related by clockwise vs. counter-clockwise orientation. Type II: 
(1,3,2,4), 
(1,3,4,2), and their counter-clockwise partners 
(1,4,2,3) and 
(1,2,4,3). For Type I, all four domain walls carry the same tension $\sigma_{\rm DW,1}$. For Type II, the two vertical walls can have different tensions from the two horizontal walls, so both $\sigma_{\rm DW,1}$ and $\sigma_{\rm DW,2}$ enter.

\begin{figure}[th!]
    \centering
   \includegraphics[width=0.6\linewidth]{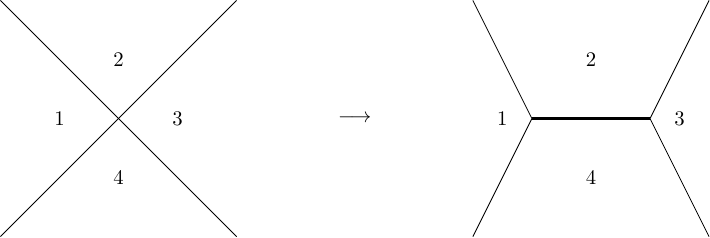}
    \caption{Illustration of a type-I X-type junction decaying into a linked Y-junction configuration when $\sigma_{\rm DW,2} < \sqrt{2}\,\sigma_{\rm DW,1}$. The thicker line in the right panel indicates a larger wall tension.}
    \label{fig:junction-decay}
\end{figure}

Whether X-type or Y-type junctions are stable depends on the ratio of the two wall tensions $\sigma_{\rm DW,1}/\sigma_{\rm DW,2}$. For example, the type-I X-junction is stable for $\sigma_{\rm DW,2} > \sqrt{2}\,\sigma_{\rm DW,1}$; otherwise, it will decay into a configuration of linked Y-type junctions to reduce the total tension-weighted length (see Fig.~\ref{fig:junction-decay} for an illustration and Ref.~\cite{PinaAvelino:2006ia} for further discussion). On the other hand, one can show that the type-II X-type junction is not stable and has a higher energy than the linked Y-type configuration. We will not consider more complicated junctions, which have been shown to be even less stable in Ref.~\cite{PinaAvelino:2006ia}.

\begin{figure}
    \centering
    \subfigure[]{\includegraphics[width=0.3\linewidth]{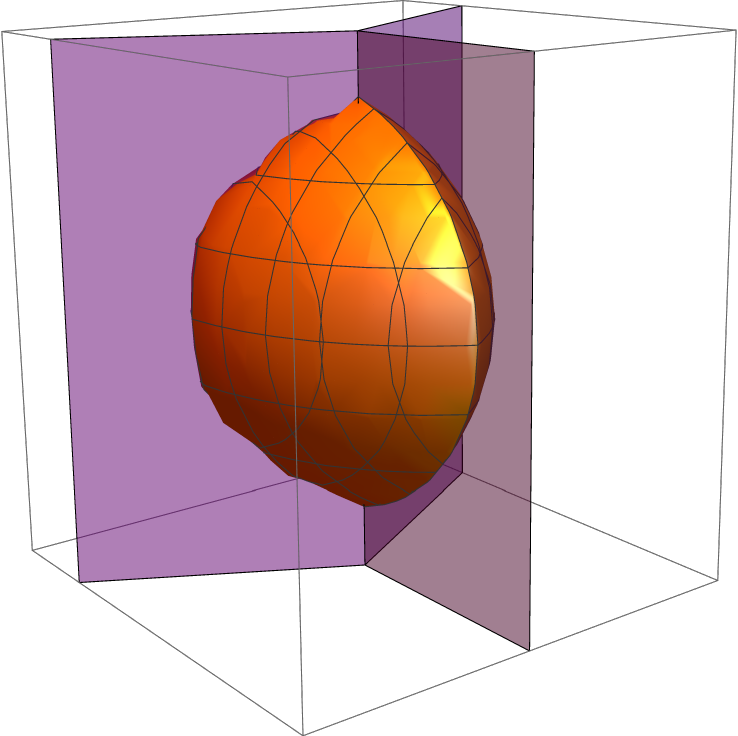}\label{fig:Z3DWJbubbleModel_sphcap}}
    \subfigure[]{\includegraphics[width=0.3\linewidth]{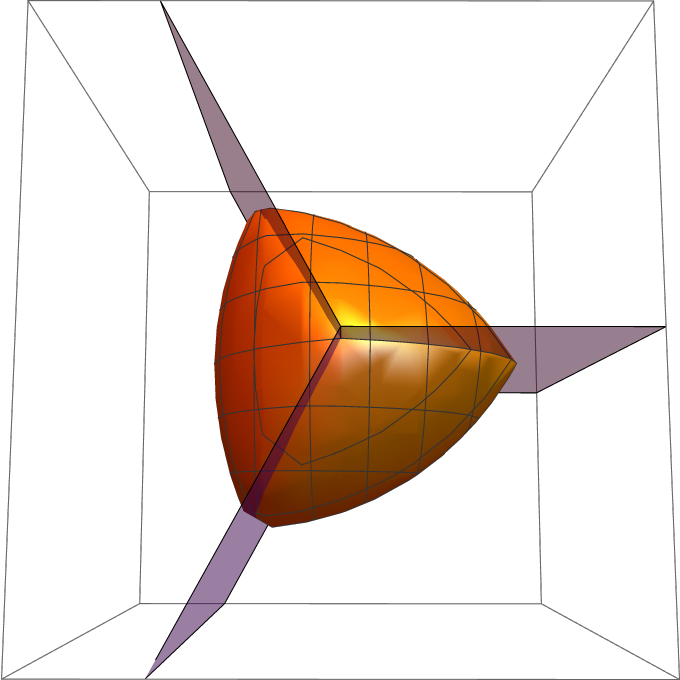}}
    \subfigure[]{\includegraphics[width=0.3\linewidth]{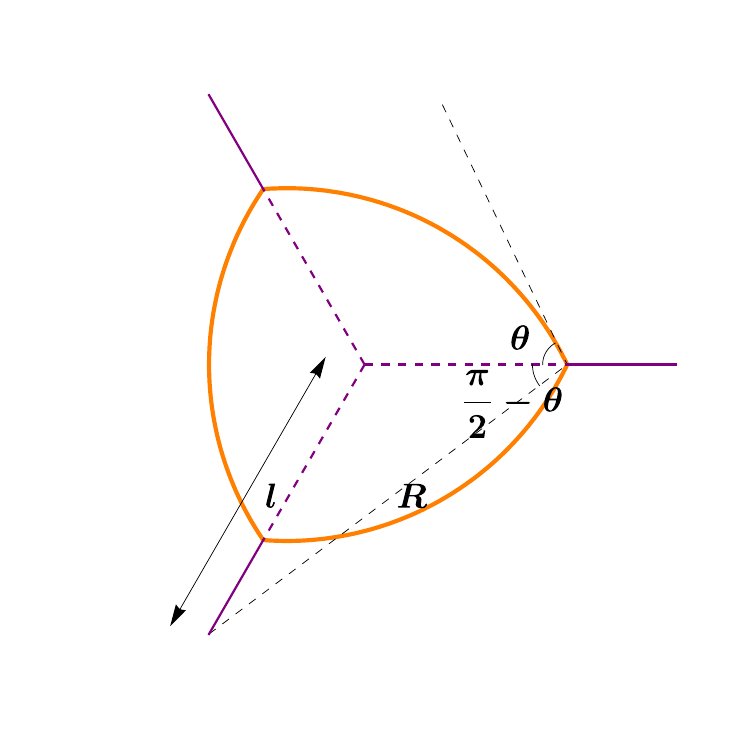}\label{fig:Z3DWJbubbleModel_sphcap_z0}}
    \caption{Illustration of the geometric shape of the spherical-cap bubble, shown in orange, together with the domain wall, shown in purple. (a) The 3d shape of the bubble on the $\mathbb{Z}_3$ junction, where the angle between adjacent walls is $2\pi/3$. To maintain continuity of the bubble across the wall, the center of the sphere is constrained to lie along the outer angle bisector of the two walls. (b) The $z=0$ slice of the bubble at the junction. (c) Notation for the $z=0$ slice. }
    \label{fig:geo_sphcap}
\end{figure}

To estimate the critical action for junction seeded bubble, we again use the thin-wall approximation and consider multiple spherical caps that are generated by the domain wall cutting a sphere. Starting with the Y-type junction with equal domain wall tension, we illustrate the geometry of the 3 sphere caps around the junction in Fig.~\ref{fig:Z3DWJbubbleModel_sphcap}. 

Denote $R$ as the radius of the sphere and $l$ as the distance between the junction center and the sphere center. The contribution of the junction region to the action scales as $R \times T_{\rm J}$, where $T_{\rm J} = \int_{r < r_0} \dd x \dd y\, \epsilon$ is the junction tension, $r_0$ is the junction size, and $\epsilon$ is the energy density. Because this contribution scales subdominantly in the limit $r_0 \ll R$, we neglect it and keep only the remaining contributions to the action, given by
\begin{align}
    &S_3^{Y-\rm J} = ~\mathcal{A}_1(R,l) \,\sigma_{\rm B} -\mathcal{V}(R,l)\, \Delta V - \mathcal{A}_2(R,l)\,\sigma_{\rm DW},\label{eqn:S3dwj_critical_thinwall} \\
    &\hspace{3mm}\mathcal{A}_1(R,l)=3\times 4 R^2\left[\arctan(\sqrt{3}\sqrt{1-x^2})-\frac{\sqrt{3}}{2}x\arctan(2\sqrt{\frac{1}{x^2}-1})\right], \nonumber \\
    &\hspace{3mm}\mathcal{V}(R,l)=3\times R^3\left[\frac{\sqrt{3}}{6}x^2\sqrt{1-x^2}+\frac{4}{3}\arctan(\sqrt{3}\sqrt{1-x^2})-\sqrt{3}x\left(1-\frac{x^2}{4}\right)\arctan(2\sqrt{\frac{1}{x^2}-1})\right], \nonumber \\
    &\hspace{3mm}\mathcal{A}_2(R,l)=\frac{3}{2}R^2\left(1-\frac{3\,x^2}{4}\right)\left[2\arctan(2\sqrt{\frac{1}{x^2}-1})-\frac{x\sqrt{1-x^2}}{1-3\,x^2/4}\right]\,,\nonumber
\end{align}
where $x = l/R$. As shown in Fig.~\ref{fig:Z3DWJbubbleModel_sphcap_z0}, the geometric relation between $\cos\theta$ and $l/R$ is
$\cos\theta = \frac{l}{R}\sin(2\pi/3)$.
Applying the Young equation~\eqref{eq:young}, $\cos\theta = \sigma_{\text{DW}}/(2\sigma_{\text{B}})$, and noting that $S_3^{Y-\rm J}$ remains a cubic function of $R$ alone, one can obtain the expression for the critical bubble action for the Y-type junction 
\begin{align}
    S_{3,c}^{Y-\rm J} = \frac{8\, \sigma_{\text{B}}^3}{3\,\Delta V^2}\biggl[&\cos^2\theta \sqrt{3-4\cos^2\theta}+3\cos\theta(\cos^2\theta-3)\arctan(\sqrt{\frac{3}{\cos^2\theta}-4})\nonumber\\
    &+6\arctan(\sqrt{3-4\cos^2\theta})\biggr]~, \qquad \mbox{for}~\, 0\leq \cos{\theta} \leq \sqrt{3}/2 ~.\label{eqn:S3c-Y120}
\end{align}
Here, one also finds $R_{c}^{Y-\rm J} = 2\sigma_{\rm B}/\Delta V$, the same critical radius as for the spherical bubbles in homogeneous nucleation. The corresponding critical action is simply the homogeneous result multiplied by the appropriate volume-reduction factor. 

As in the wall-seeded case, one can extremize the action with respect to $\cos\theta$, or equivalently $x$, to derive the Young equation as in \eqref{eqn:S3dwj_critical_thinwall}. We also note that the upper bound $\sigma_{\rm DW}/(2\sigma_{\rm B}) = \cos\theta \leq \sqrt{3}/2$ is more restrictive than in the wall-seeded case. Moreover, the contact angle $\theta$ is the same for all $z$ because the tangential planes are related by rotational symmetry; for convenience we evaluate the geometry on the $z=0$ slice in Fig.~\ref{fig:Z3DWJbubbleModel_sphcap_z0}. This reveals another advantage of this construction: inherited from the homogeneous bubble, the bubble tension $\sigma_{\text{B}}$ is uniform everywhere, which is consistent with the requirement that the tension-balance condition at the junction is identical at every point, with the same tensions and the same contact angle.

\begin{figure}[th!]
    \centering
   \includegraphics[width=0.6\linewidth]{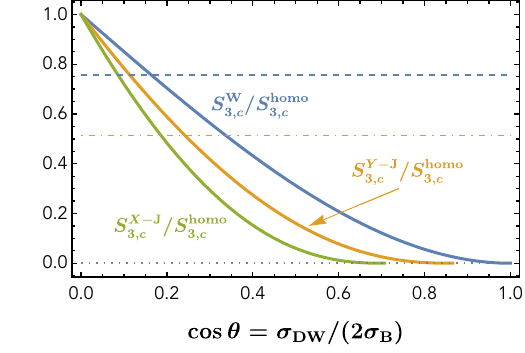}
    \caption{Ratios of the critical bubble actions for the wall-induced, Y-type–junction–induced, and X-type–junction–induced cases relative to the homogeneous spherical bubble, based on the thin-wall approximation. The orange Y-type junction curve has an upper bound at $\cos\theta = \sqrt{3}/2$, while the green X-type junction curve has an upper bound at $\cos\theta = 1/\sqrt{2}$. The horizontal blue dashed line is at approximately $3/4$ and intersects the blue curve at $\cos\theta = 0.16$. The horizontal orange dashed line is at approximately $1/2$ and intersects the orange curve at $\cos\theta = 0.24$ and the green curve at $\cos\theta = 0.19$.}
    \label{fig:action-ratio-junction}
\end{figure}

As a comparison to the homogeneous nucleation rate, we show in Fig.~\ref{fig:action-ratio-junction} the ratio of the critical‐bubble actions for the junction-induced case over the homogeneous one. One can see that the ratio decreases as $\cos{\theta}$, or equivalently the domain‐wall tension $\sigma_{\rm DW}$, increases. The junction-induced nucleation action is not only smaller than the homogeneous one, but also smaller than the domain-wall–induced one, highlighting the important role that junctions can play in first-order phase transitions.

In more complicated potentials, such as those with $\mathbb{Z}_4$ symmetry, the domain walls can have different tensions, with the ratios of these tensions depending on the specific form of the potential. Consequently, one must consider more intricate $Y$-type junction configurations. For example, a $Y$-type junction with three wall tensions $(\sigma_1, \sigma_1, \sigma_2)$ has the angles between the walls determined by Young’s equations. Denoting by $\alpha$ the angle between the two walls with equal tension $\sigma_1$, the critical action in the thin-wall approximation is given by
\begin{align}
    S_{3,c}^{Y-\rm J, II}& = \frac{4\sigma_{\rm B}^3}{\Delta V^2} \Biggl\{\frac{\cos^2\theta_1}{3\sin^2\frac{\alpha}{2}}(2\sin\alpha-\sin 2\alpha)\sqrt{1-\frac{\cos^2\theta_1}{\sin^2\frac{\alpha}{2}}}\label{eqn:S3_c_arbitaryY}\\
    &-\Biggl[4\left(\cos\theta_1-\frac{1}{3}\cos^3\theta_1\right)\arctan\left(\frac{\sqrt{\sin^2\frac{\alpha}{2}-\cos^2\theta_1}}{\cos\theta_1\cos\frac{\alpha}{2}}\right)-\frac{8}{3}\arctan\left(\frac{\sqrt{\sin^2\frac{\alpha}{2}-\cos^2\theta_1}}{\cos\frac{\alpha}{2}}\right)\nonumber\\
    &+2\left(\cos\theta_2-\frac{1}{3}\cos^3\theta_2\right)\arctan\left(\frac{\sqrt{\sin^2\alpha-\cos^2\theta_2}}{\cos\theta_2|\cos\alpha|}\right)-\frac{4}{3}\arctan\left(\frac{\sqrt{\sin^2\alpha-\cos^2\theta_2}}{\cos\theta_2|\cos\alpha|}\right)\Biggr]\Biggr\}\nonumber,
\end{align}
where $\theta_1$ is the contact angle on the walls with tension $\sigma_1$, and $\theta_2$ is the contact angle on the wall with tension $\sigma_2$ [the contact angles on the two sides of a wall are identical from the Young's equation in \eqref{eq:Young's-equation}]. Note that we have already used Young’s equation to simplify the calculation by expressing 
\begin{align}
    \frac{\sigma_1}{2\sigma_{\rm B}}=\cos\theta_1\,,\quad\frac{\sigma_2}{2\sigma_{\rm B}}=\cos\theta_2\,,\quad\frac{\sigma_2}{2\sigma_1}=\cos\frac{\alpha}{2}\,,\label{eqn:Young_arbitary_Y}
\end{align}
for $\alpha \in (\frac{\pi}{2}, \pi]$. When $\alpha = \frac{\pi}{2}$, the terms involving $\arctan$ of $\theta_2$ should be interpreted as approaching $\frac{\pi}{2}$. Note that when $\alpha = 2\pi/3$, one recovers the result in Eq.~\eqref{eqn:S3c-Y120}. When $\alpha \in [0,\frac{\pi}{2})$, one has an addition term, $(2/3)(\pi-2\alpha)\sqrt{1-\cos^2\theta_2/\sin^2\alpha}\,(2+\cos^2\theta_2/\sin^2\alpha)$, inside the curly braces.

For $X$-type junctions, we consider the simple case in which all wall tensions are equal (the unequal-tension configuration is unstable in the $\mathbb{Z}_4$ symmetry). In this case, it is straightforward to show that the angle between adjacent walls is $\pi/2$. From the geometry, one obtains 
\begin{align}
    &S_3^{X-\rm J} = ~\mathcal{A}_1(R,l)\, \sigma_{\rm B} -\mathcal{V}(R,l)\, \Delta V - \mathcal{A}_2(R,l)\,\sigma_{\rm DW}\,,\label{eqn:S3_c_Xjunction} \\
    &\hspace{3mm}\mathcal{A}_1(R,l)=4\times 4 R^2\left[\arctan(\sqrt{1-x^2})-\frac{\sqrt{2}}{2}x\arctan(\sqrt{2}\sqrt{\frac{1}{x^2}-1})\right]\,, \nonumber\\
    &\hspace{3mm}\mathcal{V}(R,l)=4\times R^3\left[\frac{1}{3}x^2\sqrt{1-x^2}+\frac{4}{3}\arctan(\sqrt{1-x^2})-\sqrt{2}x(1-\frac{x^2}{6})\arctan(\sqrt{2}\sqrt{\frac{1}{x^2}-1})\right]\,,\nonumber\\
    &\hspace{3mm}\mathcal{A}_2(R,l)=2\,R^2\left(1-\frac{x^2}{2}\right)\left[2\arctan(\sqrt{2}\sqrt{\frac{1}{x^2}-1})-\frac{\sqrt{2}x\sqrt{1-x^2}}{1-x^2/2}\right]\,, \nonumber
\end{align}
where $x=l/R$ again, and the geometry provides $\cos\theta=\frac{l}{R}\sin(3\pi/4)$. The critical action is
\begin{align}
    S_{3,c}^{X-\rm J} = \frac{32\, \sigma_{\text{B}}^3}{3\,\Delta V^2}\biggl[&\cos^2\theta \sqrt{1-2\cos^2\theta}+\cos\theta\,(\cos^2\theta-3)\arctan(\sqrt{\frac{1}{\cos^2\theta}-2})\nonumber\\
    &+2\arctan(\sqrt{1-2\cos^2\theta})\biggr]~,
\end{align}
with the action proportional to the volume-reduction factor. The ratio of $S_{3,c}^{X-\rm J}/S_{3,c}^{\rm homo}$ is shown in the green curve in Fig.~\ref{fig:action-ratio-junction}.

\section{Heterogeneous cosmological phase transition} \label{sec:Heter-Cosmo}

When one applies the heterogeneous bubble nucleation to cosmological first-order phase transition, both the temperature dependence of $S_3/T$ and the Hubble expansion play roles. One could work out some general formulas for nucleation and percolation temperatures for different dimensions of defects and under different background evolution: radiation-dominated, matter-dominated and DW-dominated universe with the equation-of-state parameter $\omega <0$. One could further compare the nucleation/percolation temperature or the so-called supercooling rate $(T_c - T_n)/T_n$ in terms of the contact angle $\theta$ (or the ratio of DW tension over the bubble wall tension).  
 
The decay rate of the false vacuum per unit of volume in space can be parametrized by 
\begin{align}
    \gamma \,=\, \frac{\Gamma}{V}=\frac{\xi\, \gamma_k\, V_k}{V}\,.
\end{align}
Here, $\Gamma$ is the decay rate; $V$ is a large 3d spatial volume; $\xi$ is the number of impurities within $V$, which is taken to be $\xi \,V_k/V = 1$ for the homogeneous case with $k=3$ and a fixed value when the impurities reach the so-called ``scaling region,'' where the number per Hubble volume becomes approximately constant; $\gamma_k$ is the decay rate per $k$-dimensional volume, having units of $1/(\mbox{s}\cdot\mbox{cm}^2)$ for domain walls with $k = 2$ and $1/(\mbox{s}\cdot\mbox{cm})$ for junctions with $k = 1$; and $V_k$ is the $k$-dimensional volume of the defects, with units of $\mbox{cm}^2$ for domain walls and $\mbox{cm}$ for junctions. Up to an overall coefficient of order unity, the decay rate $\gamma_k$ (per second per $V_k$ volume) is given by~\footnote{Note that the power of $[B(T)/(2\pi)]^{1/2}$ corresponds to the number of zero modes of the bounce solution. The bouncing profile is localized on the defect, which breaks $k$ translational symmetries along the defect, while the defect itself breaks the remaining $3-k$ translational symmetries. Therefore, the total number of zero modes remains three, and the overall prefactor continues to scale as $[B(T)/(2\pi)]^{3/2}$, just as in the homogeneous case.}
\beqa
    \gamma_k \approx T^{1+k} \left[\frac{B_k(T)}{2\pi}\right]^{3/2}\exp\left[-B_k(T)\right]~, \qquad \mbox{with}\quad B_k(T) \equiv S^{(k)}_3/T ~.
\eeqa

\subsection{Nucleation temperature $T_n$} \label{sec:nucleation-tempeature}
We first calculate the nucleation temperature, which is defined to have roughly one bubble per Hubble volume. Parametrize the relation of Hubble in temperature as $H= C\,T^\alpha$ with $\alpha = 2, \frac{3}{2}, \frac{1}{2}$ for radiation-, matter- and domain-wall-dominated universe. Note that $\alpha$ is related to the equation-of-state parameter $\omega$ as $\alpha = \frac{3}{2}(1+\omega)$. The number of bubbles in the Hubble volume is
\beqa
    \mathcal{N}&=&\xi \int_{t_c}^{t_n}\dd t\, \frac{\gamma_k}{H^{k}}\,=\,\xi \int_{T_n}^{T_c}\frac{\dd T}{T} \frac{\gamma_k}{H^{1+k}} \nonumber \\    &\approx&\xi \,\int_{T_n}^{T_c}\frac{\dd T}{T}\,\left(\frac{T}{H}\right)^{1+k}\left[\frac{B_k(T)}{2\pi }\right]^{3/2}\exp\left[-B_k(T)\right],
\eeqa
where $T_c$ is the critical temperature and $T_n$ is the nucleation temperature for the case that one true-vacuum bubble appears in each Hubble volume~\cite{Ellis:2018mja, Blasi:2022woz}. Here, we have assumed the scaling behavior for the impurities with $\xi\, V_k/V\simeq \xi \,H^{3-k}$.

For a quick phase transition with $T_c - T_n \ll T_c$, one can approximate the integration by
\beqa
    \mathcal{N}\approx 
    \xi\,\frac{ [B_k(T_n)/2\pi]^{3/2}\exp[-B_k(T_n)]}{1-(1-\alpha)(1+k)}\left(\frac{T_n}{ H}\right)^{1+k},
\eeqa
when $\alpha>1$. If one takes the condition $\mathcal{N} \approx 1$, drops the order-one factor $1 - (1-\alpha)(1+k)$, and uses $[B_k(T_n)/2\pi]^{3/2} \approx (100/2\pi)^{3/2} \approx 63$, one arrives at the critical value of $B_k(T)$ at which nucleation starts to occur:
\beqa
    B_k(T_n) \,\approx\, (1+k)\log(\frac{T_n}{H(T_n)})+\log\xi+4.1~.\label{eqn:nucleation_condition}
\eeqa

For the radiation-dominated era, the Friedmann equation implies $H = T^2/(2\epsilon M_{\rm Pl})$, where we define $\epsilon = \sqrt{45/(\pi g_\star(T))}/(4\pi)$. Here $g_\star(T)$ is the effective number of relativistic degrees of freedom, which is $106$ for SM particles when $T > \mathcal{O}(500)~{\rm GeV}$. This implies that the nucleation condition for $S_3/T$ is
\begin{align}
    B_k(T_n)=(1+k)\log(1.4\times10^{15}\times\frac{100~{\rm GeV}}{T_n})+\log\xi+4.1\,, \qquad \mbox{[radiation-dominated]}\,.
\end{align}
For the matter-dominated era, $H = H_0 \Omega_{M,0}^{1/2} (T/T_0)^{3/2}$, where $H_0$ is the current Hubble constant, $\Omega_{M,0}$ is the current matter density, and $T_0$ is the current temperature of the Universe. Therefore,
\begin{align}
    B_k(T_n)=(1+k)\log\left[1.4\times10^{22}\times\left(\frac{100~{\rm GeV}}{T_n}\right)^{\frac{1}{2}}\right]+\log\xi+4.1\,, \quad \mbox{[matter-dominated]}\,.
\end{align}
For the domain-wall–dominated era, the energy density is proportional to the temperature, $\rho = \rho_f\, T/T_f$, where the subscript $f$ denotes the formation time of the domain walls, and $\rho_f = \sigma_{\rm DW} H_f$ for domain walls with tension $\sigma_{\rm DW}$. By dimensional analysis, $\sigma_{\rm DW} \sim T_f^3$, and we assume that this era follows a radiation-dominated epoch with $H_f = T_f^2/(2\epsilon M_{\rm Pl})$. Using these relations, the Friedmann equation $H^2 = 8\pi G \rho_f\, T/(3 T_f)$ gives $H=1/\sqrt{6\epsilon}\, (T_f/M_{\rm Pl})^2\,(T M_{\rm Pl})^{1/2}$. Under these assumptions,  if we choose $T_f\approx10^6$~GeV as a reference temperature,
\begin{align}
    B_k(T_n)=&~(1+k)\log\left[1.6\times10^{16}\times\left(\frac{10^6~{\rm GeV}}{T_f}\right)^{2}\times\left(\frac{T_n}{100~{\rm GeV}}\right)^{\frac{1}{2}}\right]
    \nonumber\\ &
    ~ +\log\xi+2.8\,. \qquad\qquad  \mbox{[wall-dominated]}\,.
\end{align}
The values of $B_k(T_n)$ or $S_3/T$ for each case are summarized in Table~\ref{tab:nucleationS3T_impur_era}. Note that we take $\xi = 1$ for the radiation- and matter-dominated universes, and $\xi = 10^3$ for the domain-wall–dominated universe.

\begin{table}[h!]
    \centering
    \renewcommand{\arraystretch}{1.5}
    \begin{tabular}{c||c|c|c}
    \hline\hline
         & radiation & matter & domain wall \\ \hline
        homogeneous ($k=3$) & 144 & 208 & 159 \\ \hline
        wall-induced ($k=2$) & 109 & 157 & 122 \\ \hline
        junction/string-induced ($k=1$) & 74 & 106 & 84 \\       \hline\hline
    \end{tabular}
    \caption{The critical value of $S_3/T$ at which nucleation begins for various types of impurities that induce the phase transition. We take $\xi = 1$ for radiation- and matter-dominated universes, and $\xi = 10^3$ for the domain-wall–dominated universe.}
    \label{tab:nucleationS3T_impur_era}
\end{table}

Note that the ratios $B_k(T_n)/B_3(T_n)$ are approximately $(1+k)/4$, giving $3/4$ for wall-induced nucleation and $1/2$ for junction-induced nucleation. In Fig.~\ref{fig:action-ratio-junction}, the wall-induced nucleation rate becomes more important than the homogeneous one when $\cos\theta > 0.16$, while the $Y$-type ($X$-type) junction–induced nucleation becomes more important than the homogeneous one when $\cos\theta > 0.24$ ($0.19$). Compared to the wall-induced case, the $Y$-type ($X$-type) junction–induced nucleation is more important when $\cos\theta > 0.36$ ($0.20$).

An additional comment concerns the number $\xi$ for junctions and walls. In principle, they are related by the Euler formula and depend on the number of external walls (i.e., walls that do not terminate within the Hubble volume). One finds that the ratio $\xi_{\rm J}/\xi_{\rm DW}$ lies in the range $[1/3,\, 2/3)$ if only $Y$-type junctions are present.

\subsection{Percolation temperature $T_p$} \label{sec:percolation-tempeature}

For the percolation temperature, and depending on the dimensionality of the impurity, one has slightly different criteria for defining the percolation temperature $T_p$. For the homogeneous case, the criterion is that the false vacuum occupies $1/e$ of the total volume of the Universe, following the 3d random percolation model governed by Erd\"{o}s–R\'{e}nyi Poisson statistics~\cite{erdos59a}. For bubble percolation on 2d domain walls, the percolation threshold corresponds to an unoccupied area fraction of approximately $1-0.68=0.32$ for randomly placed overlapping discs in 2d~\cite{PhysRevB.30.3933,PhysRevE.76.051115}. In a purely 1d space, percolation occurs once the nucleated bubbles expand to occupy the entire line. We emphasize, however, that percolation in a lower-dimensional subspace does not guarantee percolation in the full 3d volume. Whether the phase transition completes depends on the number density of impurities, a condition we derive below.

For the homogeneous case, the probability of remaining in the false vacuum at a time after the critical temperature is~\cite{Guth:1981uk}
\beqa
\label{eq:p(t)-homo}
    p(t)=\exp[-\frac{4\pi}{3}\int_{t_c}^t\dd t'\, v_{\rm sh}^3\,(t-t')^3\,\gamma(t')]~.
\eeqa
Here, $t_c$ is the time at which the plasma temperature equals $T_c$, and $v_{\rm sh}$ is the bubble-wall velocity. We have ignored the scale-factor dependence since the temperature is close to $T_c$. Percolation occurs when $p(t_p) = 0.32$, which determines the percolation temperature $T_p$. 

The situation for the heterogeneous case is more complicated. First, on a lower-dimensional impurity, a percolation temperature $T_{p}^{(k)}$ can be defined using a probability analogous to Eq.~\eqref{eq:p(t)-homo}. At $T_{p}^{(k)}$, the bubbles are still localized around the impurity locations, meaning that the distances between bubbles nucleated on different impurities can remain large, depending on the impurity number density. In the diluted case, this ensemble of localized bubbles expands primarily in the transverse (co-dimension) directions, and full 3d percolation occurs only at a later time, with the final percolation temperature satisfying $T_p < T_p^{(k)}$. In the dense case, on the other hand, one has $T_p = T_p^{(k)}$, with the 3d percolation coinciding with the percolation time on the lower-dimensional impurities. 

Before calculating $T_p^{(k)}$, we first comment on how the bubbles expand prior to percolation. Based on the thin-wall approximation in Section~\ref{sec:general_nucleation}, the spherical-cap shape of the bubbles remains fixed with the same contact angle $\theta$ as they grow. A simple way to see this is by examining the extremization of the action in \eqref{eqn:S3dw_thinwall} with respect to $\cos\theta$ and $R$. Starting from the critical configuration, the action is minimized with respect to $\cos\theta$ and maximized with respect to $R$. Thus, $R$ grows while $\cos\theta$ remains unchanged as the spherical cap expands. Referring to Fig.~\ref{fig:DWbubbleModel_sphcap-c}, both $R_\parallel$ and $R_\perp$ increase while maintaining the fixed ratio $R_\parallel/R_\perp = \sin\theta /(1 - \cos\theta)$.

To estimate $T_p$ and $T_p^{(k)}$, we will also make a simplifying assumption about the wall–junction network. Specifically, we assume the presence of ``great walls" and long junctions that are parallel to each other and have Hubble-scale lengths. In general, however, the network may exhibit more intricate structures—resembling a building with ``floors" and ``ceilings"—with the detailed configuration determined by the discrete symmetry and the early-universe history.

We first estimate the lower-dimensional percolation temperature $T_p^{(k)}$. For a $k$-dimensional impurity, and following the same logic as in the 3d case in Eq.~\eqref{eq:p(t)-homo}, the probability of remaining in the false vacuum $p_{k}(t)$ in $k$ dimensions can be written as
\begin{align}
    p_{k}(t)=\exp[-\frac{2}{k}\frac{\pi^{k/2}}{\Gamma\left(\frac{k}{2}\right)}\int_{t_c}^t\dd t' \,v_{\rm sh}^{k}\,(t-t')^{k}\,\gamma_k(t')]\,.
\end{align}
Applying the same approximation as Ref.~\cite{Bai:2022kxq} and denoting $\log\gamma_k(t')\approx\log\gamma_k(t_p^{(k)})+(t'-t_p^{(k)})\beta'$, where $t_p^{(k)}$ is the percolation time and 
\begin{align}
    \beta'=&\frac{\dd\log\gamma_k(t)}{\dd t}=-\frac{TH}{\gamma_k}\frac{\dd}{\dd T}\left[T^{k+1} \left(\frac{B_k(T)}{2\pi}\right)^{3/2}\exp(-B_k(T))\right]\nonumber\\
    &=-(k+1)H-\frac{3\,TH}{2\,B_k}\frac{\dd B_k}{\dd T}+TH\frac{\dd B_k}{\dd T}\,.
\end{align}
Note that, given the approximate scaling of $B_k(T) \propto 1/(T-T_c)^2$ when $T$ is close to $T_c$, $\dd B_k/\dd T = 2B_k/(T_c - T)$ is positive and dominates over $\beta'$. One usually defines $\beta \equiv T H\,\dd B_k/\dd T|_{T=T_p^{(k)}}$, so that the probability $p_{k}(t_p^{(k)})$ that a point in the $k$-dimensional space remains outside the bubble can be approximated as
\begin{align}
    p_{k}(t_p^{(k)})\approx\exp[-\frac{2}{k}\frac{\pi^{k/2}}{\Gamma\left(\frac{k}{2}\right)}\gamma_k(t_p^{(k)})\int_{t_c}^{t_p^{(k)}}\dd t' v_{\rm sh}^{k}(t_p^{(k)}-t')^{k}e^{(t'-t_p^{(k)})\beta}]\approx\exp[-A_k\gamma_k(t_p^{(k)})\frac{1}{\beta^{k+1}}]\,,
\end{align}
with 
\begin{align}
    A_k\equiv\frac{2}{k}\frac{\pi^{k/2}\Gamma(k+1)}{\Gamma\left(\frac{k}{2}\right)}v_{\rm sh}^{k}\,.
\end{align}
The lower-dimensional percolation happens at  $p_{k}(t_p^{(k)})=0.32$, so one has 
\begin{align}
    1\approx A_k\gamma_k(t_p^{(k)})\frac{1}{\beta^{k+1}}=A_k\left(\frac{T_p^{(k)}}{\beta}\right)^{k+1}\left(\frac{B_k(T_p^{(k)})}{2\pi}\right)^{3/2}\exp[-B_k(T_p^{(k)})]\,,\label{eqn:percolation_condition1}
\end{align}
with $\beta/H = 2\,B_k\,T_p^{(k)}/(T_c - T_p^{(k)})$. The percolation condition in Eq.~\eqref{eqn:percolation_condition1} can be written as
\begin{align}
    B_k(T_p^{(k)})=(k+1)\log\left[\frac{T_p^{(k)}}{H(T_p^{(k)})}\right]-(k+1)\log\left[ \frac{2B_k(T_p^{(k)})T_p^{(k)}}{T_c - T_p^{(k)} }\right] +\log A_k +\frac{3}{2}
    \log\left[\frac{B_k(T_p^{(k)})}{2\pi}\right]\,,\label{eqn:percolation_condition2}
\end{align}
which can be used to determine the lower-dimensional percolation temperature $T_p^{(k)}$. Since the first term in the above equation provides the dominant contribution to $B_k$, one has $B_k \simeq (k+1)\,\log(M_{\rm Pl}/T_c)$.

One can also estimate the number of bubbles, $N_{\rm nuc}$, at the low-dimensional percolation time, which is 
\beqa
N_{\rm nuc} \approx \frac{1}{[H(T_p^{(k)})]^{k}}\int_{t_c}^{t}\dd t'\gamma_k(t')p_k(t')\approx\frac{1}{[H(T_p^{(k)})]^{k}}\int_{0}^{\infty}\frac{\dd \gamma_k}{\beta}e^{-A_k\gamma_k\beta^{-k-1}}=\frac{1}{A_k}\left(\frac{\beta}{H(T_p^{(k)})}\right)^{k} ~,
\eeqa
which is smaller compared to the homogeneous one for $k < 3$. The averaged bubble size at $t_p^{(k)}$ is estimate to be 
\beqa
R_p^{(k)} \simeq \frac{1}{H(T_p^{(k)})} \, N_{\rm nuc}^{1/k} ~. 
\eeqa

When the separation distance between two domain walls, approximately $\xi^{-1}/H(T_p^{(k)})$, becomes smaller than $R_p^{(k)}$, or equivalently when 
\beqa
\xi \gtrsim N_{\rm nuc}^{1/k} = A_k^{-1/k}\times \frac{\beta}{H(T_p^{(k)})} = A_k^{-1/k} \times \frac{2\,B_k\,T_p^{(k)}}{T_c - T_p^{(k)}} \equiv \xi_0 ~,
\label{eq:xi0-definition}
\eeqa
one has the 3d percolation time coinciding with the 2d percolation time, or, 
\beqa
T_p = T_p^{(k)} \,, \qquad \mbox{for}\quad \xi > \xi_0 ~. 
\eeqa
Note that $\xi_0 = \mathcal{O}(10^2\text{--}10^3)$, depending on the phase-transition temperature and the supercooling rate.

For the diluted case with $\xi < \xi_0$, the ensemble of bubbles located on the $k$-dimensional space must expand further before colliding with bubbles nucleated on a nearby impurity. The separation distance between impurities is $L(t) \approx \xi^{-1}\,[H(T_p^{(k)})]^{-1}\,(t/t_p^{(k)})^{1/\alpha}$, while the grown bubble radius is $R(t) \approx R(t_p^{(k)}) + v_{\rm sh}\,(t- t_p^{(k)})$. For $\alpha > 1$ and requiring $R(t) = L(t)$, the resulting 3d percolation temperature is 
\beqa
T_p \approx T_p^{(k)}\left( 1 - v_{\rm sh}^{-1}\,\xi^{-1}\right)\, ,
\eeqa
which is valid for $v_{\rm sh}\,\xi > \mbox{few}$. Thus, the 3d percolation temperature is generally close to the lower-dimensional percolation temperature. For the $k = 2$ domain-wall case, the ensemble of bubbles on a wall expands primarily within the plane of the wall and eventually collides with bubbles nucleated on nearby walls. For the $k = 1$ domain-wall–junction case, the ensemble of bubbles forms an effectively cylindrical configuration. These geometric differences can lead to distinct phenomenological imprints, including potentially observable consequences for gravitational waves.

\section{Domain-wall-seeded phase transition} \label{sec:domain-wall}

In this section, we use a simple two-scalar-field model to study domain-wall–seeded first-order phase transitions. Both $\mathbb{Z}_2$ and $\mathbb{Z}_n$ symmetries (focusing on the $\mathbb{Z}_3$ case) will be analyzed. We consider a cosmological scenario in which, after the first-order phase transition, all domain walls disappear due to the restoration of the discrete symmetries associated with the walls at a later stage in the evolution of the Universe. 

\subsection{$\mathbb{Z}_2$ domain walls}\label{sec:Z2}

First, consider the tree-level potential for two real scalar fields $\phi$ and $S$ with a $\mathbb{Z}_2$ symmetry associated with the $S$ field 
\begin{align}
    V_0(\phi,S) = - \frac{\mu_S^2}{2}S^2 + \frac{\lambda}{4}S^4 -\frac{\mu_\phi^2}{2}\phi^2 + \frac{\eta}{4}\phi^4  + \frac{\kappa}{2} \phi^2 S^2 \,.\label{eqn:VZ2Notemper}
\end{align}
The finite-temperature potential (assuming no additional fields coupled to the two scalar fields) is
\begin{align}
    &V_T=\frac{T^4}{2\pi^2}\left[J_B\left(\frac{M_S^2}{T^2}\right)+J_B\left(\frac{M_\phi^2}{T^2}\right)\right] \,,
\end{align}
where $J_B(y^2)=\int_0^\infty \dd x\,x^2\log[1-\mbox{exp}(-\sqrt{x^2+y^2})]$, $M_S^2 = -\mu_S^2+3\lambda\, S^2+\kappa\,\phi^2$ and $M_\phi^2 = -\mu_h^2+3\eta\,\phi^2+\kappa\,S^2$. In the high temperature limit and taking the leading two terms of the high-temperature expansion of $J_B(y^2) \approx - \frac{\pi^4}{45} + \frac{\pi^2}{12}y^2$, one has $V_T=T^2[(3\eta+\kappa)\phi^2+(3\lambda+\kappa)S^2]/24$ and the total effective potential 
\begin{align}
    &V(\phi,S,T) \equiv V_0 + V_T \approx   \frac{c_S T^2-\mu_S^2}{2}S^2 + \frac{\lambda}{4}S^4 + \frac{c_\phi T^2-\mu_\phi^2}{2}\phi^2 + \frac{\eta}{4}\phi^4 + \frac{\kappa}{2} \phi^2 S^2\,.\label{eqn:VZ2temper}
\end{align}
Here $c_\phi \equiv (3\eta + \kappa)/12$ and $c_S \equiv (3\lambda + \kappa)/12$.~\footnote{If $\phi$ is the radial mode of the SM $SU(2)$ Higgs doublet field (as considered in Ref.~\cite{Blasi:2022woz}), one has $\mu_\phi = 88~{\rm GeV}$ and $\eta = 0.129$ for the SM Higgs. In this case, $c_\phi=(2m_W^2+m_Z^2+m_h^2+2m_t^2)/(4v_{\rm EW}^2)+ \kappa/12$, $c_S=(3\lambda+4\kappa)/12$ with $v_{\rm EW}=246~{\rm GeV}$.} The high-temperature limit is not perfectly valid in this model, since the critical temperature is of the same order as $\mu_\phi$ and $\mu_S$. We nevertheless use the high-temperature effective potential here to simplify the discussion. 

For the cosmological history of this model, we consider the so-called ``two-step'' phase transition. At a higher temperature, $T_1 \equiv \sqrt{\mu_S^2/c_S}$, the vacuum expectation values (VEVs) $(\langle \phi \rangle, \langle S \rangle)$ transition from the unbroken configuration $(0, 0)$ to the $\mathbb{Z}_2$-breaking configuration $(0, \pm v_S)$, with $v_S$ being temperature dependent. At a lower temperature $T_c$, a first-order phase transition happens with the VEVs transiting from $(0, \pm v_S)$ to $(v_\phi, 0)$, with the $\mathbb{Z}_2$ symmetry restored and an additional symmetry breaking associated with the $\phi$ field. At this critical temperature, one has $V(0, \pm v_S, T_c) = V(v_\phi, 0, T_c)$.

We note that we do not address here the potentially additional $\mathbb{Z}_2$ symmetry acting on $\phi$. This symmetry may not be present if $\phi$ is interpreted as the radial mode of the SM Higgs doublet, in which case the would-be $\mathbb{Z}_2$ can be absorbed into a continuous gauge transformation. A more complete discussion of the possible breaking patterns of the $\mathbb{Z}_2 \times \mathbb{Z}_2$ model is provided in Appendix~\ref{app:differentZ2}.

As discussed in Refs.~\cite{Blasi:2022woz, Agrawal:2023cgp}, the second-step phase transition can be first order, which is the focus of our study and serves to illustrate the importance of heterogeneous phase transitions. During the first-step phase transition, the domain walls separating the $+v_S$ and $-v_S$ domains are produced via the Kibble–Zurek mechanism~\cite{Kibble:1976sj, Zurek:1985qw}. In this work, we do not follow the detailed evolution of the domain walls; instead, we simply assume that they reach the scaling regime~\cite{Leite:2011np} before the lower temperature $T_c$. The profile of the domain wall in this specific model has the known analytical expression
\begin{align}
    S_{{\rm DW}, T}(z) = v_S(T) \tanh(\frac{\lambda^{1/2}\,v_S(T)\,z}{\sqrt{2}})\,.
\end{align}
The domain wall tension is $\sigma_{\rm DW} = \int_{-\infty}^{+\infty} dz\, T_{00} = \frac{2\sqrt{2}}{3}\,\sqrt{\lambda}\,v_S^{3}(T)$.

Although it is not the main purpose of our study to analyze the model parameter space, we want to point out that one can use the Hessian operator to check the stability of the domain-wall configurations, which is present in Appendix~\ref{app:stability} and could be useful when one analyzes other models with domain-wall configuration.

The bubble nucleation rate for the second-step first-order phase transition requires finding the bounce, or saddle-point soliton, profile~\cite{Coleman:1977py, Callan:1977pt}. The rate is proportional to $e^{-S_3^{(k)}/T}$, with 
\beqa
S_3^{(k)} \equiv S_3[\mbox{soliton}]-S_3[\mbox{false vacuum}] ~,
\eeqa
which is determined by solving a set of partial differential equations (PDEs) with the appropriate boundary conditions. 

For the homogeneous phase transition, the Laplacian is $\nabla^2=\frac{\partial^2}{\partial r^2}+\frac{2}{r}\frac{\partial}{\partial r}$ in spherical coordinate, and the boundary conditions are
\begin{eqnarray}
    S(\infty) = \pm v_S(T)\,,\quad S'(0) = 0 \,, \quad \phi(\infty) = 0\,, \quad \phi'(0) = 0 ~,
\end{eqnarray}
which can be solved using the  public package like \texttt{FindBounce}~\cite{Guada:2020xnz} or \texttt{CosmoTransitions}~\cite{Wainwright:2011kj}. The bubble surface tension can be calculated in the thin-wall limit by following Ref.~\cite{Coleman:1977py} 
\beqa
    \sigma_{\rm B}=\int^{\varphi_1^{\rm F}}_{\varphi_1^{\rm *}}\dd\varphi_1~\sqrt{\sum_{i=1}^n\left(\frac{\dd\varphi_i}{\dd\varphi_1}\right)^2}\,\sqrt{2\left[V(\varphi_i)-V(\varphi_i^{\rm *})\right]}\,.\label{eq:bubble_tension_field_thinwall}
\eeqa
Here we use $\varphi_i$ to denote the multiple fields $(S, \phi)$ in this model; $\varphi_i^{\rm F}$ denotes those at the false vacuum; $\varphi_i^{\rm *}$ represents the field values at the side of the true vacuum with $V(\varphi_i^{\rm *}) = V(\varphi_i^{\rm F})$. In Appendix~\ref{app:sigma_DW-sigma_B}, we show that $\sigma_{\rm DW} < 2\sigma_{\rm B}$, so the contact angle in this model satisfies $\cos\theta < 1$. 

For the wall-seeded case, we use cylindrical coordinates, with the $z$ direction perpendicular to the wall and $\rho$ the radial distance along the wall. The Laplacian is $\nabla^2=\frac{\partial^2}{\partial \rho^2}+\frac{1}{\rho}\frac{\partial}{\partial \rho}+\frac{\partial^2}{\partial z^2}$ and the boundary conditions are
\begin{eqnarray}
  &&   S(\rho,\pm \infty) = \pm v_S(T) \,,\quad  S(\infty, z) = S_{{\rm DW},T}(z)\,,\quad \frac{\partial S(0, z)}{\partial \rho} = 0 \,, \nonumber \\    
  &&  \phi(\rho,\pm \infty) = 0\,,\quad \phi(\infty,z) = 0\,, \quad \frac{\partial \phi(0, z)}{\partial \rho} = 0 \,.  
\end{eqnarray}
To solve the PDEs exactly, we use an algorithm based on the mountain-pass theorem (MPT)~\cite{AMBROSETTI1973349} (see Refs.~\cite{Agrawal:2022hnf, Agrawal:2023cgp} for a detailed description). For the $\mathbb{Z}_2$ domain-wall case, our results obtained using MPT agree very well with those in Refs.~\cite{Agrawal:2022hnf, Agrawal:2023cgp}. 

Aside from using the MPT algorithm to obtain the exact solution (up to numerical precision), we also employ two approximate methods to compute $S_3^{(k)}$. The first is based on the thin-wall approximation discussed around Eq.~\eqref{eqn:S3dw_thinwall}, using the bubble wall tension $\sigma_{\rm B}$ from the homogeneous case, as pointed out in Ref.~\cite{Blasi:2022woz}.~\footnote{The contact angle $\cos\theta = \sigma_{\text{DW}}/(2\sigma_{\text{B}})$ approaches a constant as $T \to T_c$, because $\sigma_{\text{DW}}$ is largely insensitive to $T_c$ and $\sigma_{\text{B}}$ also becomes constant in this limit. In the language of critical exponents, $\cos\theta$ is scale independent.} 

The second approximate method is named as ``$\tanh$-profile", where we introduce the following hyperbolic tangent function to model the field profile along the sphere cap
\begin{align}
    \varphi(r)=\frac{v_{\varphi,\text{F}}-v_{\varphi,\text{T}}}{2}\left(1+\tanh\left[\frac{r-r_{\varphi}}{\sqrt{2}\,d_{\varphi}}\right]\right)+v_{\varphi,\text{T}}\,.
\end{align}
Here, $r$ is the radial distance from the center of the bubble, $d_{\varphi}$ represents the wall thickness, and $r_{\varphi}$ denotes the transition radius. The quantities $v_{\varphi,\text{F}}$ and $v_{\varphi,\text{T}}$ are the VEVs of $\varphi$ in the false and true vacua, respectively. We have verified that this profile matches the bubble profile for the homogeneous phase transition very well.

The action evaluated on these field profiles becomes a function of the variational parameters, including $d_\varphi$, $r_\varphi$, and geometric parameters such as the contact angle. Extremizing the action with respect to these parameters yields an approximate critical configuration and the corresponding action. We also note that one may use the homogeneous bubble profile obtained from \texttt{FindBounce} as the initial configuration for the variational procedure to accelerate convergence. 

\begin{figure}[th!]
    \centering
    \includegraphics[width=0.48\linewidth]{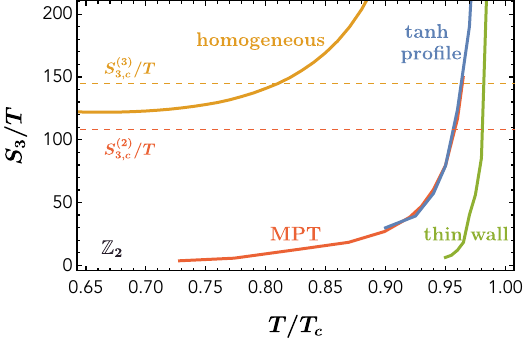}\hspace{3mm}
    \includegraphics[width=0.48\linewidth]{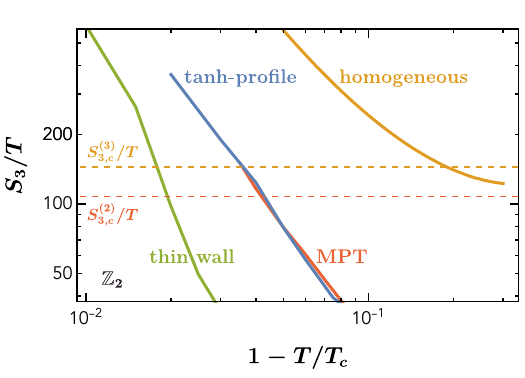}
    \caption{Left panel: The action $S_3/T$ as a function of $T$ for the model with $\mathbb{Z}_2$ domain walls. The model parameters are $\kappa=1.3$, $\lambda = 1.6$, $\eta = 0.129$, $\mu_\phi=88$~GeV and $\mu_S = 127$~GeV.  For the wall-seeded actions, we include the results from the exact solution based on the mountain-pass theorem (MPT), the thin-wall approximation, and the tanh-profile approximation. For this model point, $T_c = 109~\text{GeV}$ and $\cos\theta = \sigma_{\rm DW}/(2\sigma_{\rm B}) = 0.72$ as $T \to T_c$. The horizontal dashed lines indicate the critical nucleation actions, with $S^{(3)}_{3,c}/T = 144$ for homogeneous nucleation and $S^{(3)}_{3,c}/T = 109$ for wall-seeded nucleation. Right panel: Same as the left panel, but zoomed in on the region with $T$ close to $T_c$. } 
    \label{fig:Z2DW_S3T_plot}
\end{figure}

In Fig.~\ref{fig:Z2DW_S3T_plot}, we choose a representative model point with $\kappa = 1.3$, $\lambda = 1.6$, $\eta = 0.129$, $\mu_\phi = 88~\text{GeV}$, and $\mu_S = 127~\text{GeV}$ (the same example used in Ref.~\cite{Agrawal:2022hnf} and has a stable domain wall via Appendix~\ref{app:stability}) to show the action as a function of $T$ using different methods. For the wall-seeded nucleation, one can see that the tanh-profile method agrees very well with the exact MPT result, which justifies our later application of the tanh-profile approximation to the junction-seeded case. The thin-wall approximation underestimates the action.

The nucleation temperature—given by the intersection with the red horizontal line at $S^{(2)}_{3,c}/T = 109$—is $T_n^{(2)} = 0.95\,T_c$. For comparison, the action from homogeneous nucleation intersects the orange horizontal line at $S^{(3)}_{3,c}/T = 144$, yielding a nucleation temperature of $T_n^{(3)} = 0.81\,T_c$, which is lower than the wall-seeded case. Therefore, heterogeneous nucleation seeded by domain walls is more important than homogeneous nucleation for this model. We also note that the contact angle is $\cos\theta = 0.72$, which is sufficiently large for the wall-seeded nucleation to dominate, as indicated in Fig.~\ref{fig:action-ratio-junction}. 

\subsection{$\mathbb{Z}_{n\geq 3}$ domain walls}\label{sec:Z3}

As a representative model similar to the $\mathbb{Z}_2$ case, we introduce a complex scalar field $S$ that transforms under the discrete $\mathbb{Z}_n$ symmetry with $n \geq 3$, together with a real scalar field $\phi$. The general tree-level potential is
\beqa
    V_0(S, \phi) &=& - \mu_0^2\,S^* S\,+\,\lambda_1 (S^*S)^2 \,-\, \lambda_2\,\mu_0^{4-n}\,(S^n\,+\,S^{*n}) \nonumber \\
    && -\,\frac{m^2}{2}\,\phi^2\,+\,\frac{\eta_3\,m}{3}\,\phi^3\,+\,\frac{\eta_4}{4}\,\phi^4\,+\,\kappa\, \phi^2\,S^*S \,.\label{eqn:Z3potential-S}
\eeqa
For $n \leq 4$, the tree-level potential is renormalizable. For $n > 4$, the operator $S^n$ becomes non-renormalizable, implying that $\lambda_2 \ll 1$ and that the theory possesses an approximate $U(1)_S$ symmetry. In this study, we restrict ourselves to the $n = 3$ case with a $\mathbb{Z}_3$ discrete symmetry.

For later convenience, we choose a linear parametrization of the complex $S$ field in terms of its real and imaginary components as $S = (s + i\,a)/\sqrt{2}$. The zero-temperature potential then becomes 
\beqa
V(s,a,\phi,T=0) &=& -\frac{\mu_0^2}{2}\,(s^2+a^2) \,+\,\frac{\lambda_1}{4}\,(s^2+a^2)^2 \,-\, \frac{\lambda_2\,\mu_0^{4-n}}{2^{(4-n)/2}}\sum_{l<(n-2)/2} (-1)^l\,C_{n}^{2l}\,s^{n-2l}\,a^{2l}\nonumber\\
    &&-\frac{m^2}{2}\,\phi^2\,+\,\frac{\eta_3\,m}{3}\,\phi^3\,+\,\frac{\eta_4}{4}\,\phi^4\,+\,\frac{\kappa}{2}\, \phi^2 \,(s^2+a^2) \,.\label{eqn:Z3potentialT=0}
\eeqa

The temperature correction is calculated to be 
\beqa
    V_T=\frac{T^4}{2\pi^2}\,\left[J_B\left(\frac{M_s^2}{T^2}\right)+J_B\left(\frac{M_a^2}{T^2}\right)+J_B\left(\frac{M_\phi^2}{T^2}\right)\right]\,,
\eeqa
where
\begin{align}
    M_s^2 =& -\mu_0^2+\lambda_1\, (3s^2+a^2)+\frac{\lambda_2\,\mu_0^{4-n}}{2^{(4-n)/2}}\,n(n-1)\sum_{l<(n-2)/2}(-1)^lC_{n-2}^{2l}\,s^{n-2l-2}\,a^{2l}+\kappa\,\phi^2\,,\\
    M_a^2 =& -\mu_0^2+\lambda_1\, (3a^2+s^2)+\frac{\lambda_2\,\mu_0^{4-n}}{2^{(4-n)/2}}\,n(n-1)\sum_{l<(n-2)/2}(-1)^l\,C_{n-2}^{2l-2}\,s^{n-2l}\,a^{2l-2}+\kappa\,\phi^2\,,\\
    M_\phi^2 =& -m^2+2\eta_3\,m\,\phi + 3\eta_4\,\phi^2+\kappa\,(s^2+a^2) \,.
\end{align}
In the high temperature limit and using the expansion of $J_B(y^2) \approx - \frac{\pi^4}{45} + \frac{\pi^2}{12} y^2$ again, one has~\footnote{For $n = 3$, there is an additional correction to the term $\lambda_2\,\mu_0 (S^3 + S^{*3})$ in the high-temperature limit, which is small for perturbative couplings. }
\beqa
V(S, \phi, T) \,\equiv\, V_0 + V_T  &\approx&  -(\mu_0^2-c_1 T^2)\,S^* S+\lambda_1\, (S^*S)^2- \lambda_2\,\mu_0^{4-n}\,(S^n+S^{*n}) \nonumber\\
    &&-\frac{m^2-c_2T^2}{2}\,\phi^2+\frac{\eta_3\,m}{3}\,\phi^3+\frac{\eta_4}{4}\,\phi^4+\kappa\,\phi^2\, S^*S \,,
\eeqa
with $c_1= (4\lambda_1+\kappa)/12$ and $c_2=(3\eta_4+2\kappa)/12$.

To simplify our presentation, we define the dimensionless fields and parameters $\bar{s}=\sqrt{\lambda_1}s/\mu_0$, $\bar{a}=\sqrt{\lambda_1}a/\mu_0$, $\bar{\phi}=\sqrt{\lambda_1}\phi/\mu_0$, $\bar{T}=\sqrt{\lambda_1}T/\mu_0$, $\bar{\eta}_3=\eta_3\,m/(\mu_0 \sqrt{\lambda_1})$, $\bar{\eta}_4=\eta_4/\lambda_1$, $\bar{\kappa}=\kappa/\lambda_1$, and normalize the coordinates as $\bar{z}=\mu_0\,z$ and $\bar{\rho}=\mu_0\,\rho$ for cylindrical coordinates. We also set $\eta_3 = 0$ to reduce the number of free parameters, corresponding to an additional $\mathbb{Z}_2$ symmetry acting on $\phi$ that can be gauged away. The critical temperature $T_c$ for the first-order phase transition from the false vacuum to the true vacuum can be calculated by equating the values of the two potentials   
\beqa
    V(T)|_{\left(0, v_s(T)\,e^{2\pi i/3}\right)} &=&-\frac{\mu_0^4}{\lambda_1}\frac{1}{12}\left(\chi+\sqrt{\chi^2+1-c_1\bar{T}^2}\right)^3\left(3\sqrt{\chi^2+1-c_1\bar{T}^2}-\chi\right)\,, \\
    V(T)|_{\left(v_\phi(T),0\right)}&=&-\frac{\mu_0^4}{\lambda_1}\frac{1}{4\bar{\eta}_4}\left[\left(\frac{m}{\mu_0}\right)^2- c_2 \bar{T}^2\right]\,,
\eeqa
with $\chi\equiv 3\lambda_2/\sqrt{8\lambda_1}$.

\begin{figure}[th!]
    \centering
    \includegraphics[width=0.31\linewidth]{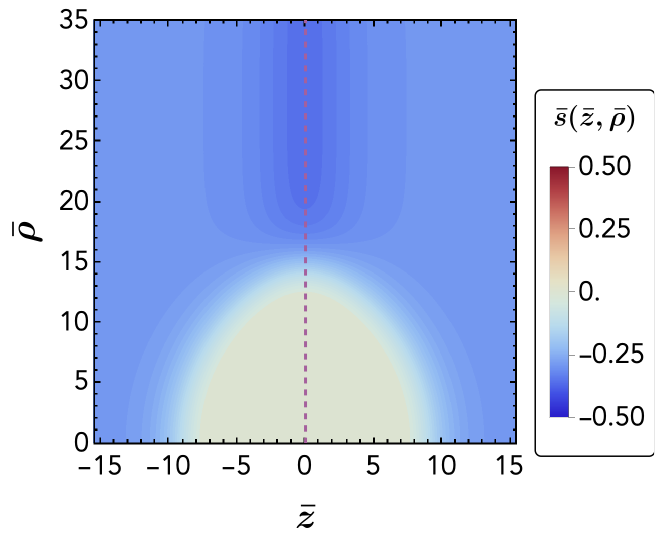}
    \includegraphics[width=0.31\linewidth]{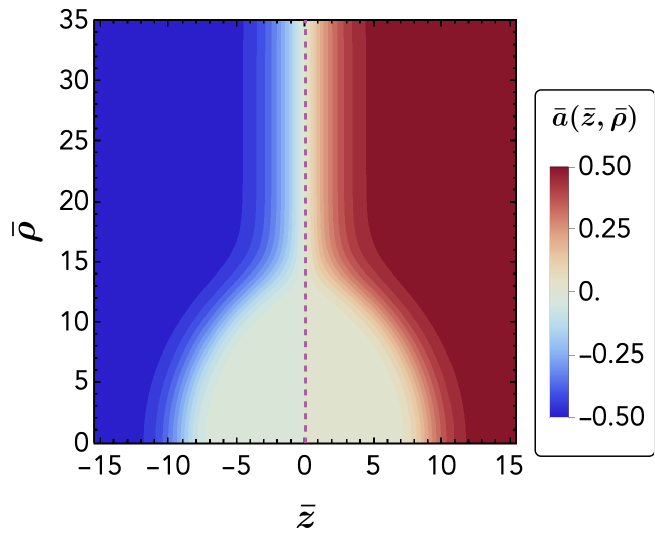}
    \includegraphics[width=0.31\linewidth]{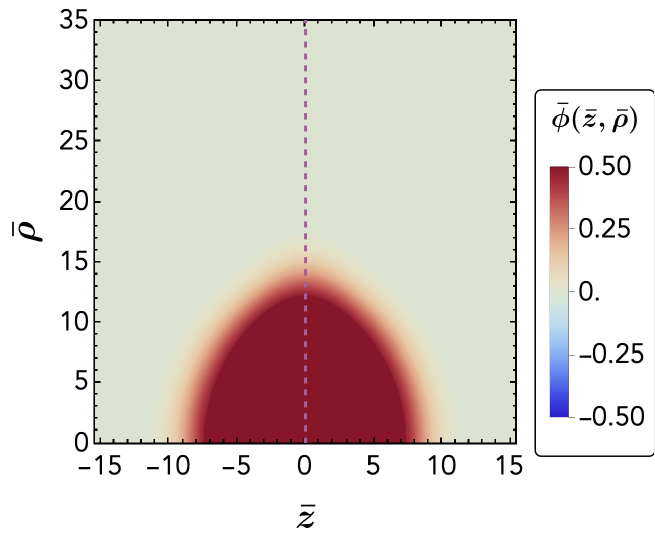}
    \caption{The field profiles—shown in the left, middle, and right panels for the $(\bar{s}, \bar{a}, \bar{\phi})$ fields, respectively—correspond to the wall-seeded bubble configuration around the $\mathbb{Z}_3$ domain wall (indicated by the purple dashed line) and are obtained using the MPT algorithm. The model parameters are $\chi = 0.1$, $m/\mu_0 = 1.2$, $\lambda_1 = 1$, $\bar{\kappa} = 7.5$, $\bar{\eta}_4 = 0.6$, and $\eta_3 = 0$, with $T = 0.92\, T_c$ and $T_c = 0.974\, \mu_0$. }
    \label{fig:Z3DW_bubbleprofile}
\end{figure}

The cosmological history of this $\mathbb{Z}_3$ model is similar to that of the previous $\mathbb{Z}_2$ model, featuring a first-order phase transition at the critical temperature $T_c$, during which the vacuum transitions from $(\langle \phi \rangle, \langle S \rangle) = (0, v_S(T)\,e^{2\pi l i /3})$ with $l = 0, 1, 2$ to the vacuum $(v_\phi(T),0)$. For temperatures above $T_c$ but below $T_1 \equiv \mu_0/\sqrt{c_1}$, we assume that domain walls have already been produced in the early Universe. The domain-wall profiles connecting regions with different $l$ can be obtained numerically (see also Refs.~\cite{Wu:2022stu, Wu:2022tpe} for domain-wall solutions of $S$ without $\phi$ at zero temperature). Unlike the $\mathbb{Z}_2$ case, where $\langle S \rangle = 0$ at the center of the wall and the symmetry is restored there, the center of a $\mathbb{Z}_3$ wall still has a nonzero value of $\langle S \rangle$ (see the region away from the bubble in Fig.~\ref{fig:Z3DW_bubbleprofile}), determined by extremizing the action. 

As a demonstration of the wall-seeded bubble configuration, we choose parameter values $m = 600~{\rm GeV}$, $\mu_0 = 500~{\rm GeV}$, $\chi = 0.1$, $\bar{\kappa} = 7.5$, and $\bar{\eta}_4 = 0.6$, which yield stable domain walls according to the Hessian-operator analysis in Appendix~\ref{app:stability}. The critical temperature is calculated to be $T_c = 0.974\,\mu_0$. We show the half-bubble field configurations in Fig.~\ref{fig:Z3DW_bubbleprofile} for a representative temperature $T = 0.92\,T_c$, obtained using the MPT algorithm.

In this figure, the vertical purple line marks the location of the original domain wall, with the perpendicular direction corresponding to $\bar{z}$ and the cylindrical radial coordinate given by $\bar{\rho}$. For the two fields $\bar{s}$ and $\bar{a}$ comprising the complex field $S$, the bubble center has field values approximately equal to zero (up to an exponentially suppressed factor of $\exp(-r_S/d_S)$, where $r_S$ is the bubble radius and $d_S$ the bubble thickness), indicating restoration of the discrete symmetry inside the bubble. Meanwhile, the field $\bar{\phi}$ develops a nonzero value inside the bubble, as shown in the right panel of Fig.~\ref{fig:Z3DW_bubbleprofile}.

The action as a function of temperature for the wall-seeded nucleation is shown in Fig.~\ref{fig:Z3DW+DWJ_eg1}, using both the MPT algorithm and the tanh-profile approximation. In the tanh-profile approach, we take the wall thicknesses to be equal, $d_s = d_a = d_\phi \equiv d$, and use a common radius for the $s$ and $a$ fields, $r_s = r_a \equiv r_S$, while allowing an independent radius $r_\phi$ for the $\phi$ field. The tanh-profile approximation reproduces the exact MPT solution very well, remaining within 20\% for temperatures close to $T_c$. 


\section{Junction-seeded phase transition} \label{sec:junction}

As discussed in Section~\ref{sec:general_nucleation}, domain-wall junctions exist for $\mathbb{Z}_n$ symmetries with $n \ge 3$. In this section, we demonstrate that junction-seeded nucleation can be more important not only than homogeneous nucleation, but also than wall-seeded nucleation. For $\mathbb{Z}_3$, the simplest $Y$-type junction has an angle of $2\pi/3$ between adjacent walls, which serves as the illustrative example in our study. We also note that for more complicated $\mathbb{Z}_n$ symmetries, the domain-wall network can exhibit different tessellation patterns of the plane, due to the various relative tensions and ratios of walls connecting different vacua; see Ref.~\cite{Saffin:1999au} for a detailed discussion.

\begin{figure}[th!]
    \centering
    \includegraphics[width=0.45\linewidth]{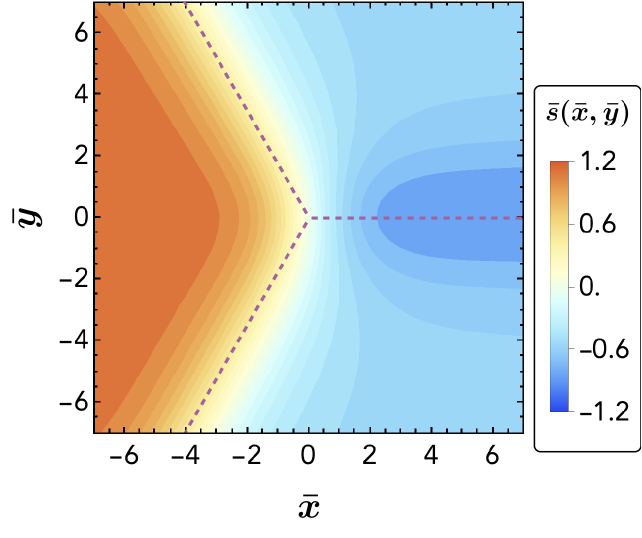}
    \includegraphics[width=0.45\linewidth]{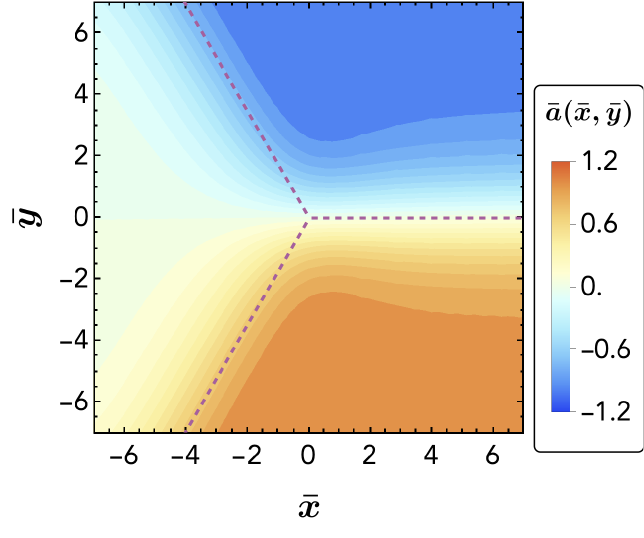} \vspace{3mm} \\
    \includegraphics[width=0.45\linewidth]{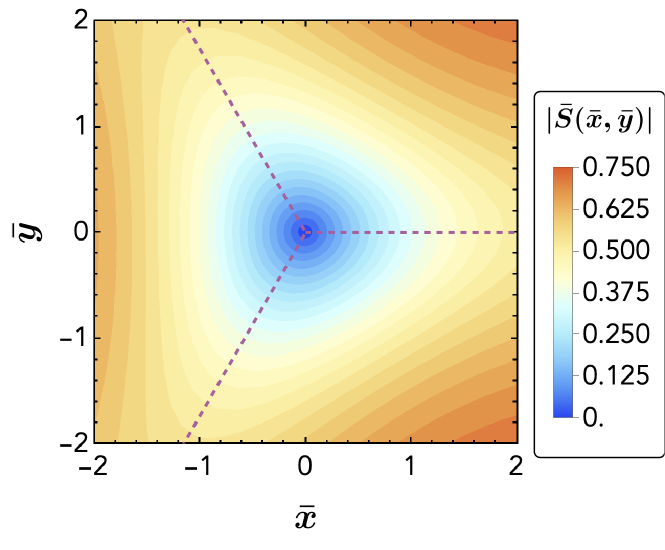}
    \includegraphics[width=0.47\linewidth]{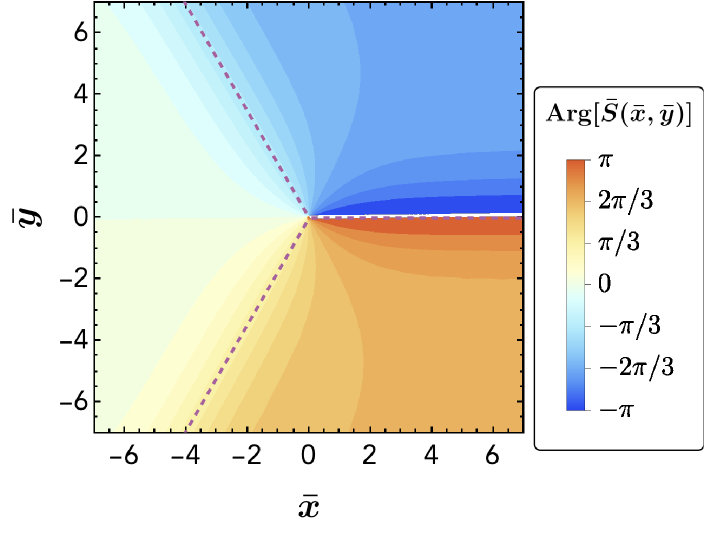}
    \caption{Example field profiles around a $Y$-type junction along the $z$ direction, based on the $\mathbb{Z}_3$ model in \eqref{eqn:Z3potentialT=0} with $T = 0$ and $\chi \equiv 3\lambda_2/\sqrt{8\lambda_1} = 0.1$. The purple dashed lines denote the locations of the domain walls. The upper two panels correspond to the two linear fields, while the lower two panels correspond to the radial and phase fields.}
    \label{fig:DWJ_contour_example}
\end{figure}

In Fig.~\ref{fig:DWJ_contour_example}, we show an example of the field profiles for the $Y$-type junction of the $\mathbb{Z}_3$ model at $T = 0$ and $\chi = 0.1$. Both linear and angular parametrizations are presented in this plot. As seen from the lower-left panel of Fig.~\ref{fig:DWJ_contour_example}, the junction size is of order the domain-wall thickness, and the core of the junction has $S = 0$, indicating restoration of the discrete symmetry. Away from the junction region, the centers of the walls exhibit nonzero but reduced values of $|S|$, corresponding to partial symmetry restoration. One also notes that in the limit $\chi \rightarrow 0$, the discrete symmetry is enlarged to a continuous $U(1)$ symmetry, and the junction effectively becomes a string (see Ref.~\cite{Blasi:2024mtc} for string-seeded phase transition).  

\begin{figure}[th!]
    \centering
    \includegraphics[width=0.31\linewidth]{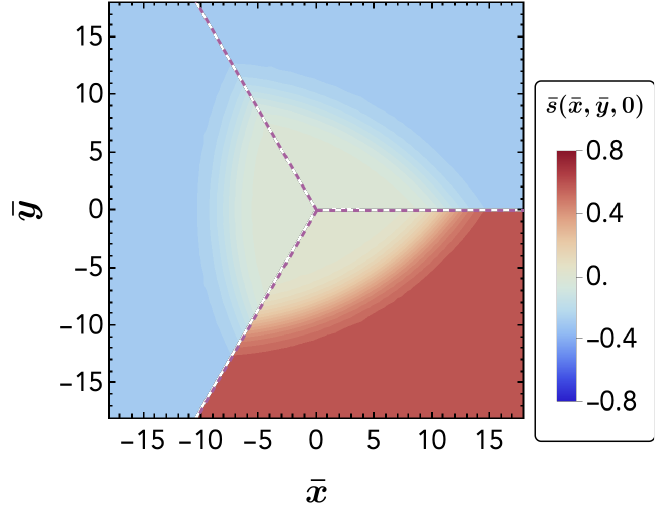}
    \includegraphics[width=0.31\linewidth]{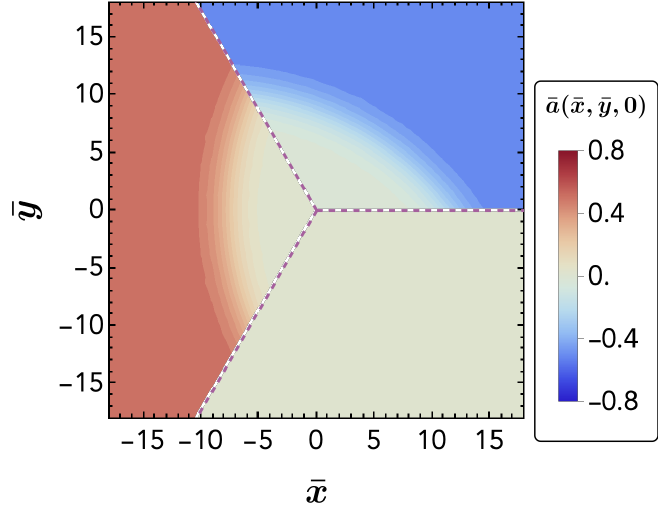}
    \includegraphics[width=0.31\linewidth]{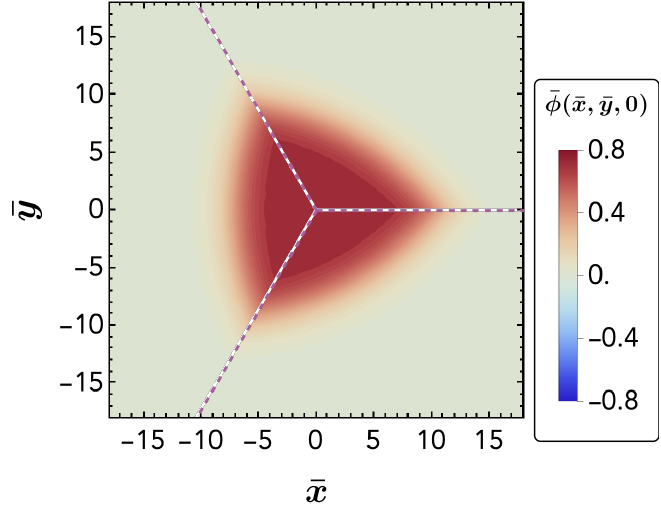} \vspace{3mm} \\
    \includegraphics[width=0.31\linewidth]{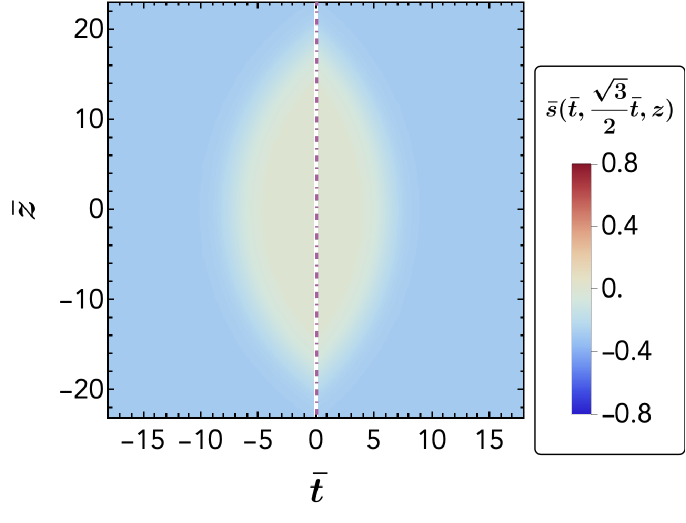}
    \includegraphics[width=0.31\linewidth]{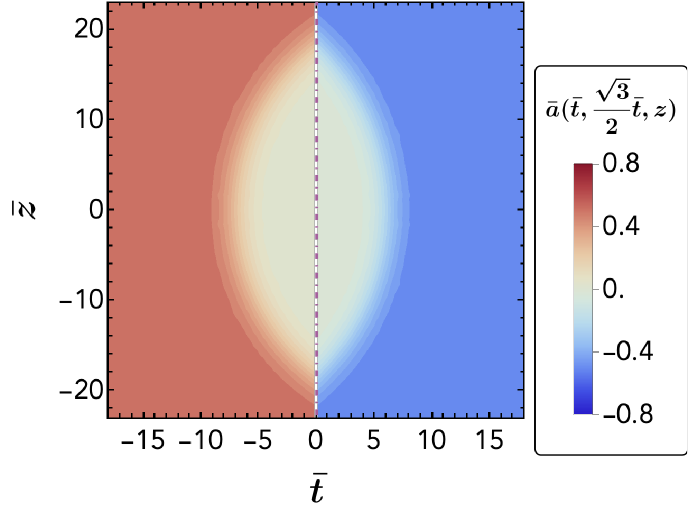}
    \includegraphics[width=0.31\linewidth]{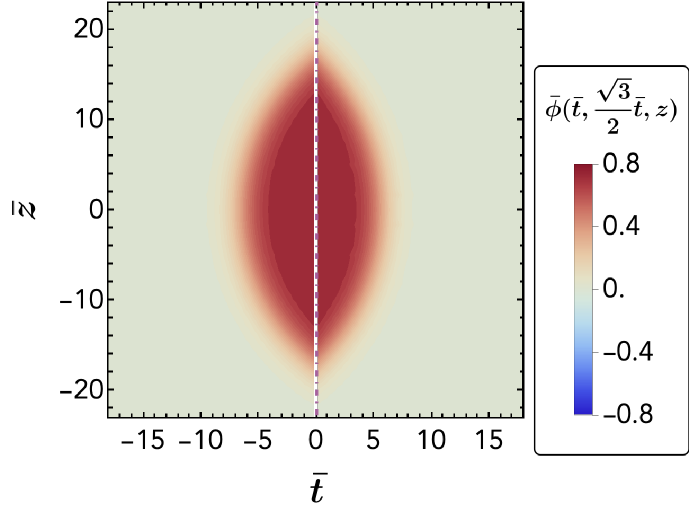}
    \caption{Top panels: Field profiles for $s$ (left), $a$ (middle), and $\phi$ (right) in the $x$–$y$ plane for the $Y$-type-junction-seeded critical bubble configuration, obtained using the tanh-profile approximation. The model parameters are the same as in Fig.~\ref{fig:Z3DW_bubbleprofile}. The purple dashed lines denote the locations of the domain walls. Lower panels: Same as the upper panels, but shown in the plane spanned by $z$ and the direction $y = \frac{\sqrt{3}}{2} x$. The purple dot-dashed line denotes the location of the wall junction. The detailed wall and junction structures outside the bubble region (as shown in Fig.~\ref{fig:DWJ_contour_example}) are not displayed here.}
    \label{fig:Z3DWJ_bubbleprofile_tanh}
\end{figure}

For the $Y$-type junction–seeded bubble configuration calculated using the tanh-profile method, we show an example in Fig.~\ref{fig:Z3DWJ_bubbleprofile_tanh}, displaying both the $x$–$y$ plane (upper panels) and a representative slice in the $z$ direction along $y = \frac{\sqrt{3}}{2} x$ (lower panels). The model parameters are the same as in Fig.~\ref{fig:Z3DW_bubbleprofile}. Similar to the wall-seeded bubble configuration, the center of the bubble has $|S| \approx 0$, indicating restoration of the discrete symmetry. Comparing the two right panels with the right panel of Fig.~\ref{fig:Z3DW_bubbleprofile}, one finds that the junction–seeded bubble occupies a smaller volume than the wall-seeded bubble, highlighting the enhanced importance of the junction–seeded configuration.

\begin{figure}[t!]
    \centering
    \includegraphics[width=0.45\linewidth]{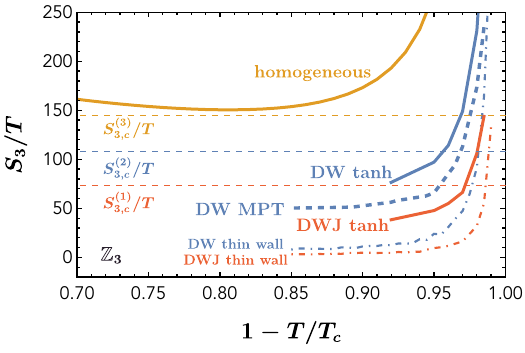}\hspace{3mm}
    \includegraphics[width=0.45\linewidth]{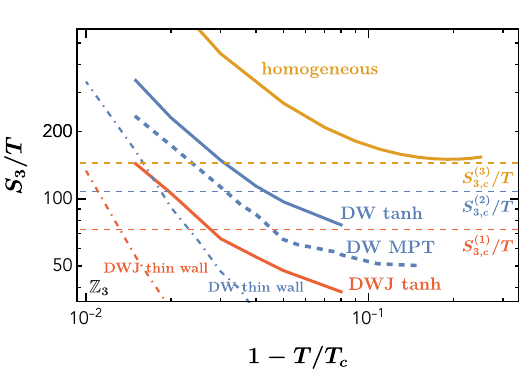}
    \caption{Left panel: The action $S_3/T$ as a function of $T$ for the model with $\mathbb{Z}_3$ domain walls. The model parameters are the same as in Fig.~\ref{fig:Z3DW_bubbleprofile}. For the wall-seeded actions (denoted as DW), we include the results from the exact solution based on the mountain-pass theorem (MPT), the thin-wall approximation, and the tanh-profile approximation. For the domain-wall-junction-seeded actions (denoted as DWJ), we include results from both the tanh-profile and thin-wall approximations. For this model point, the contact angle satisfies $\cos\theta = \sigma_{\rm DW}/(2\sigma_{\rm B}) = 0.61$ as $T \to T_c$. The horizontal dashed lines indicate the critical nucleation actions: $S^{(3)}_{3,c}/T = 144$ for homogeneous nucleation, $S^{(3)}_{3,c}/T = 109$ for wall-seeded nucleation, and $S^{(2)}_{3,c}/T = 74$ for junction-seeded nucleation. Right panel: Same as the left panel, but zoomed in on the region with $T$ close to $T_c$.}
    \label{fig:Z3DW+DWJ_eg1}
\end{figure}

In Fig.~\ref{fig:Z3DW+DWJ_eg1}, we show the action $S_3/T$ as a function of $T$ for a representative model point and for all three possible nucleation channels: homogeneous, domain-wall–seeded, and domain-wall–junction–seeded. For the domain-wall–seeded case, we present results obtained from the MPT algorithm, the tanh-profile approximation, and the thin-wall approximation. For the domain-wall–junction–seeded case, due to the numerical complexity of implementing the MPT algorithm, we show only the tanh-profile and thin-wall results.

For this particular model point, the $S_3/T$ curve for homogeneous nucleation does not intersect the critical value $S^{(3)}_{3,c}/T$, indicating that homogeneous nucleation cannot complete the phase transition. In contrast, both wall-seeded and junction-seeded nucleation do complete the transition. Using the tanh-profile results, the nucleation temperatures are $T_n^{(2)} = 0.957\,T_c$ for wall-seeded nucleation and $T_n^{(1)} = 0.973\,T_c$ for junction-seeded nucleation, clearly demonstrating that junction-seeded nucleation is more efficient.

We also note that the action obtained using the tanh-profile approximation is larger than that from the MPT algorithm in the wall-seeded case ($T_n^{(2)} = 0.969\,T_c$ using the MPT results). The exact action for the junction-seeded case is similarly expected to be smaller than the tanh-profile result. Therefore, the true nucleation temperature for the junction-seeded case should be close to—but slightly below—the quoted value for $T_n^{(1)}$.

In this section, we have used a representative model point to illustrate one of the main conclusions of our study: heterogeneous nucleation involving domain walls and junctions can play an important role. For any specific model, however, a detailed calculation is required to determine whether homogeneous or heterogeneous nucleation dominates the cosmological phase transition. On the other hand, we emphasize that intuition from ordinary nucleation theory—through the contact angle and the thin-wall approximation—provides useful guidance for estimating the relative importance of different nucleation channels, as demonstrated in Fig.~\ref{fig:action-ratio-junction}. 

\section{Discussion and conclusions}\label{sec:conclusions}

The heterogeneous phase transition leads to interesting phenomenological consequences in cosmology. The gravitational waves generated by such a first-order phase transition differ substantially from those of a homogeneous transition. Even disregarding the new parameter space opened by bubbles nucleating on defects—regions that are inaccessible to homogeneous transitions—the resulting GW spectrum is qualitatively distinct. After the bubbles reach terminal velocity, sound waves dominate GW production, as discussed in Ref.~\cite{Blasi:2024mtc} and simulated in Ref.~\cite{Blasi:2023rqi}. Although our interpretation of percolation in the dense domain-wall regime [$\xi \gg \xi_0$ with $\xi_0$ defined in \eqref{eq:xi0-definition}] differs from theirs, the main phenomenological conclusions remain similar. For dilute defects, sound waves are generated only after percolation along the defects has completed. Planar bubbles, formed by the merging of defect-seeded bubbles, then expand and interact with the surrounding plasma. The amplitude of the scaling part of the GW spectrum scales as $(\beta/H)^{-1}$, where $\beta^{-1}$ characterizes the transition timescale, and the peak frequency scales with $\beta$. Since $\beta/H \sim \xi$, a larger $\xi$ yields a smaller GW amplitude and a higher peak frequency. Junction-induced first-order phase transitions can produce yet another class of GW spectra, because the plasma interacts with cylindrical bubbles that percolate along the junctions. A full exploration of this phenomenon requires detailed simulations of the dynamics of domain-wall and junction networks.

Another source of gravitational waves is bubble collisions, which can be non-negligible depending on the model. This contribution is particularly important for dense domain walls, since the phase transition completes as soon as bubbles percolate on the sub-dimensional impurity network, and the sound-wave contribution is further suppressed because the bubbles interact predominantly with the defects rather than the plasma. The amplitude from bubble collisions scales as $(\beta/H)^{-2}$, while the peak frequency is again set by $\beta$. At the $k$-dimensional percolation temperature $T_p^{(k)}$ with $k<3$, one has, $\beta/H| _{T_p^{(k)}}=2B_k T_p^{(k)}/(T_c-T_p^{(k)})$. Although $B_k$ is smaller than its homogeneous counterpart, the fact that $T_p^{(k)}$ lies closer to $T_c$ than the homogeneous percolation temperature $T_p$ (if the latter exists) leads to a competition between these two effects, and the resulting GW amplitude may be larger or smaller than in the homogeneous case depending on the model. In the dilute-defect regime, there are two distinct percolation times, each associated with bubble collisions. Percolation on defects at $T_p^{(k)}$ gives $\beta/H| _{T_p^{(k)}}$, while true 3d percolation at $T_p$ yields $\beta/H|_{T_p} \sim \xi$. Since the redshift factor satisfies $T_p / T_p^{(k)} \approx 1 - v_{\rm sh}^{-1}\xi^{-1} \approx 1$, the relative locations of the two GW peaks are essentially determined by the ratio of the corresponding $\beta/H$ values. Consequently, the combined spectrum should exhibit two peaks: a higher-frequency peak with smaller amplitude from percolation at $T_p^{(k)}$, and a lower-frequency peak with larger amplitude from percolation at $T_p$.

The heterogeneous phase transition also has implications for electroweak baryogenesis when one identifies $\phi$ as the radial mode of the SM Higgs doublet. Inside the heterogeneous bubble corresponding to the broken phase (true vacuum), the Higgs field still reaches its VEV at the relevant temperature; consequently, the sphaleron rate is exponentially suppressed. The overall picture remains the same as in the homogeneous case: the bubble walls interact with different fermion modes in distinct ways, just as they do for a spherical bubble. However, whether the transition is sufficiently strong to prevent washout of the generated baryon asymmetry depends on the specific model details. We leave a detailed exploration of this possibility to future work. 

In conclusion, we have investigated defect-induced first-order phase transitions—heterogeneous cosmological phase transitions—in contrast to the traditional homogeneous case. We provided a physically intuitive description of bubbles nucleated on impurities using spherical geometries truncated by defects, with contact angles determined by Young’s equation. We discussed the general bubble shapes and critical actions for domain walls as well as for $Y$- and $X$-type junctions. We then computed the nucleation and percolation temperatures for heterogeneous cosmological phase transitions, confirming that defect-seeded bubbles nucleate more efficiently than homogeneous ones. Following this general framework, we used $\mathbb{Z}_2$- and $\mathbb{Z}_3$-symmetric potentials as field-theory examples to study first-order phase transitions seeded by domain walls and junctions. To better understand the bounce profiles, we employed the mountain-pass theorem algorithm to obtain numerical solutions, and constructed approximate descriptions using both the $\tanh$-profile and the thin-wall spherical-cap models. All methods consistently demonstrate that heterogeneous transitions can proceed faster than homogeneous ones, opening a new avenue for studying cosmological phase transitions and their phenomenological consequences. 

\vspace{0.2cm}
\subsubsection*{Acknowledgments}

We would like to thank Haotian Cao, Michael Nee, Subhojit Roy and Isaac Wang for helpful discussion. This work is supported by the U.S. Department
of Energy under the contract DE-SC-0017647 and DEAC02-06CH11357 at Argonne National
Laboratory.


\appendix

\section{Different breaking patterns in $\mathbb{Z}_2\times\mathbb{Z}_2$ model}\label{app:differentZ2}

In this appendix, we list more soliton configurations of $\mathbb{Z}_2\times\mathbb{Z}_2$ model, which could be studied independently in the future:
\begin{itemize}
    \item[(1)] The profile of $\phi$ is always even, and the $\phi$ field takes its nonzero VEV everywhere. Depending on the position, there are two cases:
    \begin{itemize}
        \item[(a)] When the $S$ domain wall disappears, this corresponds to the case discussed in Section~\ref{sec:Z2} and represents the best-understood breaking pattern.
        \item[(b)] The $S$ domain wall is not restored. We denote this configuration as $(v_{\phi 2}, \pm v_{S 2})$. Since $\phi$ asymptotically approaches $v_{\phi 2}$, the equation of motion for $\phi$ forces $S^2$ to become constant as well, and the VEVs can then be computed from
        \begin{align}
            \begin{cases}
                \eta\, v_{\phi2}^2 +\kappa \,v_{S2}^2=\mu_\phi^2-c_\phi\, T^2\\
                \kappa\, v_{\phi2}^2 +\lambda\, v_{S2}^2=\mu_S^2-c_S\, T^2
            \end{cases}\Rightarrow
            \begin{cases}
                v_{\phi2}^2=\frac{\kappa\,\mu_S^2-\lambda\,\mu_\phi^2-(\kappa\, c_S-\lambda\, c_\phi)T^2}{\kappa^2-\lambda\,\eta}\\
                v_{S2}^2=\frac{\kappa\,\mu_\phi^2-\eta\,\mu_S^2-(\kappa\, c_\phi-\eta\, c_S)T^2}{\kappa^2-\lambda\,\eta}
            \end{cases}\,.\label{eqn:vevs_1b_phi_S}
        \end{align}
        These values correspond to a local maximum of the potential. Ref.~\cite{Azzola:2024pzq} studies electroweak baryogenesis from domain walls with this type of configuration.
    \end{itemize} 
    \item[(2)] The profile of $\phi$ is always even, and a nontrivial VEV of $\phi$ develops inside the $S$ domain wall, while asymptotically it approaches zero. We denote this configuration as $(v_{\phi 3}, \pm v_{S3})$, where $v_{\phi 3}$ is the value of $\phi$ at the center of the wall. The requirement that $\phi$ vanishes at spatial infinity similar to case (a) leads to  
    \begin{align}
        -\mu_\phi^2+c_\phi T^2+\kappa\frac{\mu_S^2-c_S T^2}{\lambda}>0 \,.\label{eqn:req_phi=0atfar}
    \end{align} 
    \item[(3)] $\phi$ develops a domain wall, and $S$ may or may not have a nontrivial profile. If $S$ also develops a domain wall, then both $\mathbb{Z}_2$ symmetries are ultimately broken. This can occur, for example, when $\kappa = 0$, which decouples the two fields so that each can form its own domain wall independently. We denote this situation as case (3a). If instead $S$ has even parity, then the preserved $\mathbb{Z}_2$ symmetry is initially associated with $\phi$ but is later carried by $S$, and we denote this as case (3b).
\end{itemize}

We will determine which configurations can serve as the initial or final vacua of a phase transition using the Hessian-operator analysis reviewed in Appendix~\ref{app:stability}. It turns out that only cases (1a), (2), and (3b) admit stable domain-wall solutions.

\section{The stability of the soliton solutions}\label{app:stability}

For both the vacuum solutions and the instanton solution, one must solve the classical equations of motion obtained from the first-order variation of the Euclidean action,
\begin{align}
    I(\phi,S) = \int \dd^D x \left[\frac{1}{2}\partial_\mu \phi \partial^\mu \phi +\frac{1}{2}\partial_\mu S\partial^\mu S + V(\phi,S)\right]\,,\label{eqn:formal_Euc_aciton}
\end{align}
which implies
\begin{eqnarray}
    \nabla^2 \varphi = \frac{\partial}{\partial \varphi}V(\phi,S),
\end{eqnarray}
where $\varphi = \phi, S$, with the boundary condition that the fields approach either the false or the true vacuum at spatial infinity. Since we focus on thermal tunneling in this paper, the indices $\mu = 1, 2, 3$ carry a positive metric so that the Euclidean action coincides with the Hamiltonian. To determine whether a solution of the equations of motion is stable under perturbations in field space, one needs the Hessian operator of the action. Taking Eq.~\eqref{eqn:formal_Euc_aciton} as an example, where the potential depends explicitly only on the fields and not on their derivatives, we vary the classical solution $(\phi_0, S_0)$ to $(\phi_0 + \delta\phi,, S_0 + \delta S)$ and obtain 
\begin{align}
    \delta I(\phi, S)\supset \int \dd^D x\frac{1}{2}\left[\begin{pmatrix}
        \delta\phi & \delta S
    \end{pmatrix}\hat{O}(\phi_0,S_0)\begin{pmatrix}
        \delta\phi \\ \delta S
    \end{pmatrix}\right]\,,
\end{align}
where the Hessian operator for this action 
\begin{align}
    \hat{O}(\phi_0,S_0)=\eval{\begin{pmatrix}
    -\partial_\mu\partial^\mu +\frac{\partial^2 V}{\partial \phi^2} &\frac{\partial^2 V}{\partial \phi\,\partial S}\\
    \frac{\partial^2 V}{\partial \phi\,\partial S} & -\partial_\mu\partial^\mu +\frac{\partial^2 V}{\partial S^2}
    \end{pmatrix}}_{\phi=\phi_0,\, S=S_0}\, .
\end{align}
If $\hat{O}$ is semi–positive definite, the solution $(\phi_0, S_0)$ is a local minimum and corresponds to a vacuum, either true or false. If negative eigenvalues exist, the solution is unstable and represents a saddle point~\cite{Coleman:1977py, Callan:1977pt}. This Hessian formalism can be readily generalized to configurations with more fields or to more complicated Lagrangians that include derivative couplings.

In the model discussed in Refs.~\cite{Blasi:2022woz, Agrawal:2023cgp}, as in Section~\ref{sec:Z2}, the prerequisite for a two-step phase transition is that the domain wall of $S$ remains stable when the Higgs field is included in the theory. Intuitively, one might expect that, along the entire domain-wall profile, the Higgs field must always have a positive quadratic coefficient so that no tachyonic mode appears. However, this requirement is overly strong, as we demonstrate below. The necessary and sufficient condition for the $S$ domain wall to represent a vacuum configuration is that the action be a local minimum in Hilbert space.

Note that in the high-temperature limit, the finite-temperature corrections modify only the quadratic coefficients, so we use the temperature-independent potential in Eq.~\eqref{eqn:VZ2Notemper} to examine the possible vacuum configurations listed in Appendix~\ref{app:differentZ2}. Later, the temperature-dependent effects can be incorporated straightforwardly by using the effective quadratic coefficient. In case (1a), we have $S_0 = S_{\text{DW}}(z)$ and $\phi_0 = 0$. The corresponding eigenvalue problem is given by
\begin{align}
    &\frac{\dd^2}{\dd z^2}f_n +\left[-\mu_\phi^2+\kappa\, v_S^2 \tanh^2\left(\frac{\lambda^{1/2}\,v_S\,z}{\sqrt{2}}\right)\right]f_n = e_{f,n} f_n\,,\label{eqn:eigenfun_hh}\\
    &\frac{\dd^2}{\dd z^2}g_m +\left[-\mu_S^2+3\eta\, v_S^2 \tanh^2\left(\frac{\lambda^{1/2}\,v_S\,z}{\sqrt{2}}\right)\right]g_m = e_{g,m} g_m\,.\label{eqn:eigenfun_SS}
\end{align}
They can be solved analytically by P\"{o}sch-Teller potential. For Eq.~\eqref{eqn:eigenfun_hh}, one has,
\begin{align}
    \frac{e_{0,f}+\mu_\phi^2-\kappa v_S^2}{\lambda v_S^2}=-\frac{\iota^2}{2} \,,
\end{align}
for the smallest eigenvalue $e_{0,f}$ with $\frac{\kappa}{\lambda}=\frac{\iota(\iota+1)}{2}$. By requiring $e_{0,f}>0$, we have
\begin{align}
  \frac{1}{4}\sqrt{1+8\frac{\kappa}{\lambda}}-\frac{1}{4}-\frac{\mu_\phi^2}{\mu_S^2}>0~~~ \text{i.e.}~~~ \kappa > \lambda\left( \frac{\mu_\phi^2}{\mu_S^2} + 2\,\frac{\mu_\phi^4}{\mu_S^4} \right) 
  ~.\label{eqn:stableZ2_general} 
\end{align}
One can also verify that the smallest eigenvalue of Eq.~\eqref{eqn:eigenfun_SS} is always zero, which demonstrates the stability of the domain wall itself in the absence of additional fields. This zero eigenvalue corresponds to the breaking of translational invariance along the $z$ direction due to the presence of the wall. This establishes the necessary and sufficient condition for the $S$ domain wall to represent a vacuum configuration.

Building on the Hessian-operator analysis, we can identify the parameter region that allows other possible soliton solutions and examine their stability. Note that Eq.~\eqref{eqn:stableZ2_general} is a stronger requirement than demanding the Higgs VEV to vanish at infinity or imposing $\kappa > \lambda\, \mu_h^2 / \mu_S^2$. In the parameter space lying between these two conditions, one can in principle obtain a nontrivial Higgs profile, corresponding to case (1b).

\begin{figure}[th!]
    \centering
    \includegraphics[width=0.43\linewidth]{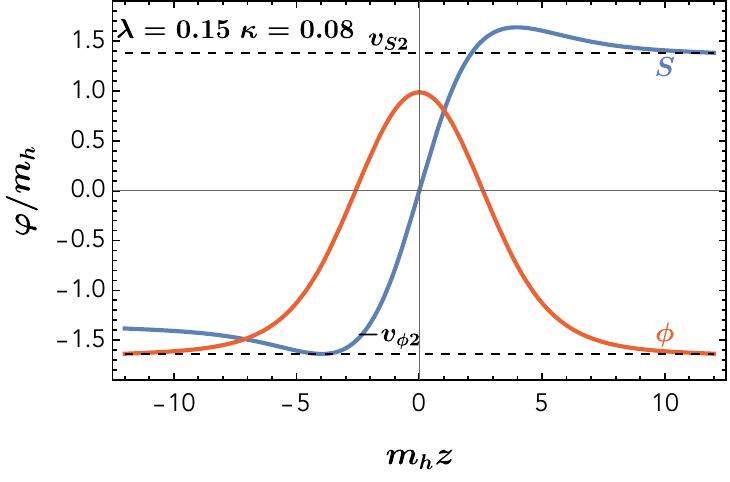}
    \includegraphics[width=0.43\linewidth]{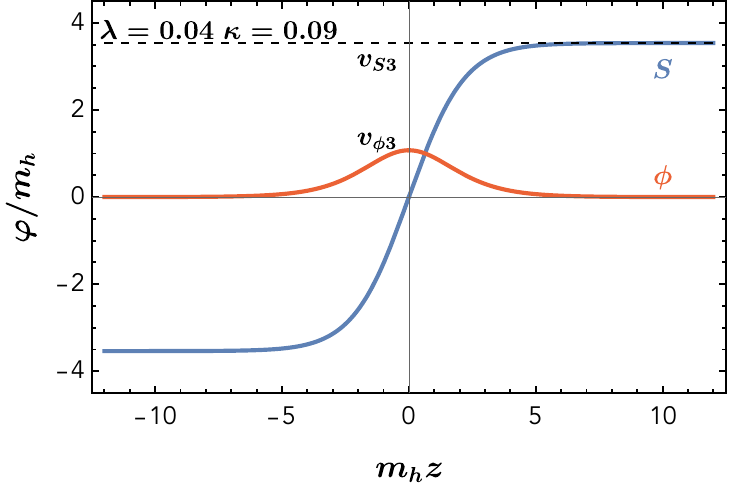} \vspace{3mm}\\
    \includegraphics[width=0.43\linewidth]{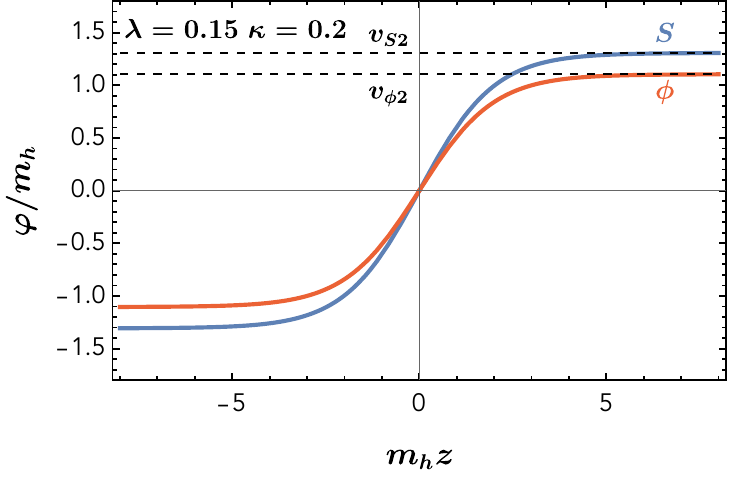}
    \includegraphics[width=0.43\linewidth]{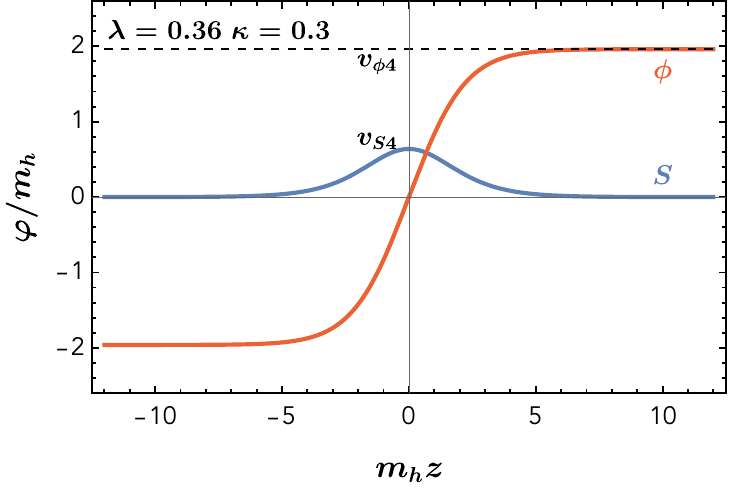}
    \caption{Additional soliton configurations under the $\mathbb{Z}_2 \times \mathbb{Z}_2$ potential of Eq.~\eqref{eqn:VZ2Notemper}, with $\eta = 0.13$, $\mu_\phi = m_h/\sqrt{2}$, and $\mu_S = \mu_\phi$. The examples correspond to cases (1b), (2), (3a), and (3b), arranged from top to bottom and left to right respectively.}
    \label{fig:Z2_soliton_manycases}
\end{figure}

It is instructive to apply the Hessian-operator analysis to the other configurations listed in Appendix~\ref{app:differentZ2}. As an example, we take the parameters of $\phi$ to match those of the electroweak Higgs field, namely $\eta = 0.13$ and $\mu_\phi = m_h/\sqrt{2}$, and we fix $\mu_S = \mu_\phi$. The resulting profiles are shown in Fig.~\ref{fig:Z2_soliton_manycases}, corresponding to cases (1b), (2), (3a), and (3b), arranged from top to bottom and left to right. After solving the equations of motion, we evaluate the Hessian operator $\hat{O}$ on each solution. Since both $\phi$ and $S$ develop nontrivial profiles, the Hessian operator is not diagonal, and a numerical evaluation is required. We find that only cases (2) and (3b) possess positive-definite Hessians, whereas cases (1b) and (3a) each exhibit at least one negative eigenvalue for the profiles displayed in Fig.~\ref{fig:Z2_soliton_manycases}. In addition, all four configurations contain a zero mode associated with the breaking of translational invariance, as in the single–domain-wall case. A similar classification has been carried out for the two-Higgs-doublet model in Ref.~\cite{Battye:2025uny}. 

There is one additional condition that can allow an $S$-field domain-wall solution together with a nontrivial Higgs profile. At $z = 0$, the domain wall satisfies $S(z = 0) = 0$. The Higgs profile must obey $\phi'(z = 0) = 0$, so it is natural for the Higgs to take the value $\phi(z = 0) = v_\phi = \sqrt{\mu_\phi^2/\eta}$. This Higgs VEV feeds back into the effective mass of the $S$ field at $z = 0$, giving $- \mu_S^2 + \kappa \,v_\phi^2 = - \mu_S^2 + \kappa\,\mu_\phi^2/\eta$. To keep this quantity negative, one requires $\kappa < \eta\,\frac{\mu_S^2}{\mu_\phi^2}$, although this condition is sufficient but not necessary. On the other hand, the opposite inequality, $\kappa > \eta\,\frac{\mu_S^2}{\mu_\phi^2}$ is required for case (1b) and (3). 

One can generalize this method to other topological defects in order to determine whether a given ``vacuum" configuration is truly a vacuum. It is important to emphasize that in any model, one should first verify the positive definiteness of the Hessian to ensure that the configuration represents a well-defined vacuum corresponding to a local minimum. If the Hessian is block-diagonal, one block will correspond to the intrinsic stability of the defect itself, as in Eq.~\eqref{eqn:eigenfun_SS}, while the remaining blocks must be checked explicitly for positive eigenvalues. This strategy has been applied, for example, in Ref.~\cite{Blasi:2024mtc} in the context of string-seeded phase transitions.

One can apply the same reasoning in the presence of finite temperature. In case (1a), it is only necessary to ensure that the domain-wall solution remains stable at the critical temperature $T_c=\sqrt{\frac{\mu_\phi^2-\sqrt{\eta/\lambda}\,\mu_S^2}{c_h-\sqrt{\eta/\lambda}\,c_S}}$, for the potential including the high-temperature corrections in Eq.~\eqref{eqn:VZ2temper}. This value of $T_c$ is a distinctive feature of case (1), and such a temperature does not necessarily exist in other cases. The stability condition of Eq.~\eqref{eqn:stableZ2_general} then becomes
\begin{align}
\left(\sqrt{\frac{\eta}{\lambda}}+\frac{1}{4}-\frac{1}{4}\sqrt{1+8\frac{\kappa}{\lambda}}\right)\left(\frac{\mu_\phi^2}{c_\phi}-\frac{\mu_S^2}{c_S}\right)\left(c_\phi-c_S\sqrt{\frac{\eta}{\lambda}}\right)>0 ~.
\label{eq:stablity-Hessian-9}
\end{align}
However, one should examine this condition more carefully. Physically, we require the following (i) $T=0$ the Higgs vacuum with nonzero VEV must be the global minimum, which implies $-\mu_\phi^4/\eta<-\mu_S^4/\lambda$; (ii) The critical temperature $T_c$ must exist; (iii) $T=T_c$, $v_S^2(T)=(\mu_S^2-c_S T^2)/\lambda>0$, otherwise no domain-wall solution is present. From the first two requirements, one obtains $c_\phi-c_S\sqrt{\eta/\lambda}>0$, and combining all three gives $\mu_\phi^2/c_\phi-\mu_S^2/c_S<0$. The signs of the latter two parentheses in \eqref{eq:stablity-Hessian-9} in the stability condition are therefore fixed by physical considerations, leaving only the first parenthesis as nontrivial. One can verify that this parenthesis is always negative when $\mu_\phi^2/c_\phi-\mu_S^2/c_S<0$ in case (1a), implying that the $S$ domain-wall solution remains stable in the presence of the Higgs field as long as the domain wall itself exists. 

Note that (iii) is as weak as we can set. When $T>T_c$, $v_S^2$ will more tend to be negative, and at least we hope $T=T_c$ the domain wall can be formed instantly.  On the other hand, $\frac{1}{4}\sqrt{1+8\kappa/\lambda}-\frac{1}{4}-(\mu_\phi^2-c_\phi T^2)/(\mu_S^2-c_S T^2)<0$ gives the largest temperature for the domain wall is unstable in Hilbert space with respect to additional field,
\begin{align}
    T_a^2=\frac{\mu_\phi^2-\mu_S^2\left(\frac{1}{4}\sqrt{1+8\frac{\kappa}{\lambda}}-\frac{1}{4}\right)}{c_\phi-c_S\left(\frac{1}{4}\sqrt{1+8\frac{\kappa}{\lambda}}-\frac{1}{4}\right)}.
\end{align}
For cases (2) and (3b), if we want a nontrivial phase transition to occur, we require that $T_a$ be larger than $T_c$ so that the system can begin in the configurations $(v_{\phi 2}, \pm v_{\phi 2})$ or $(v_{\phi 3}, \pm v_{\phi 3})$. The condition $T_a > T_c$ is equivalent to 
\begin{align}
    \sqrt{\frac{\eta}{\lambda}}+\frac{1}{4}-\frac{1}{4}\sqrt{1+8\frac{\kappa}{\lambda}}>0\,,~\text{i.e.}~\frac{\kappa}{\lambda}<2\frac{\eta}{\lambda}+\sqrt{\frac{\eta}{\lambda}}\,,\label{eqn:SDW_unstable_Tc}
\end{align}
which is the same condition as requiring the domain wall to become unstable immediately above $T_c$.

\section{Constraint on $\sigma_{\rm DW}/\sigma_{\rm B}$ for the scalar field model}
\label{app:sigma_DW-sigma_B}

From the setup of the domain wall and the homogeneous bubble, one can show that $\sigma_{\text{DW}} < 2\sigma_{\text{B}}$ generally holds under the thin-wall approximation. As discussed in Appendix B of Ref.~\cite{Agrawal:2023cgp}, the bubble-wall tension is 
\begin{align}
    \sigma_{\rm B}=\int_{R-\delta R}^{R+\delta R}\dd r \left(\frac{1}{2}\sum_i \left[\frac{\dd \varphi_i}{\dd r}\right]^2+\left[V(\phi_i)-V(\varphi_i^*)\right]\right)\,,
    \label{eq:sigma-B-in-r}
\end{align}
where $\varphi_i$ denote all the fields in the potential and $\varphi_i^*$ is the release point on the true-vacuum side satisfying $V(\varphi_i^{\rm *}) = V(\varphi_i^{\rm F})$. In thin wall limit, $V(\varphi_i^{\rm T}) \approx V(\varphi_i^{\rm F})$, so the release point is close to the true vacuum and the derivative terms in Eq.~\eqref{eq:sigma-B-in-r} is 0. Using the equations of motion, one can derive Eq.~\eqref{eq:bubble_tension_field_thinwall}. In the following proof, we will use the fact $\varphi_i^{\rm T}=\varphi_i^{\rm *}$.

We take $\varphi_1$ to be the field $a$ and consider the bubble transiting from $\varphi_1(R - \delta R) = a^{\rm T}$ to $\varphi_1(R + \delta R) = a^{\rm F_1}$, where we denote false vacuum $\ev{S}^{\rm F_1}=v_S(T)\,e^{2\pi i /3}$ and $\ev{S}^{\rm F_2}=v_S(T)\,e^{4\pi i /3}$. The bubble surface tension can then be written as
\begin{align}
    \sigma_{\rm B}=\int^{\varphi_1^{\rm F_1}}_{\varphi_1^{\rm T}=0}\dd\varphi_1~\sqrt{\sum_{i=1}^n\left(\frac{\dd\varphi_i}{\dd\varphi_1}\right)^2}\sqrt{2\left[V(\varphi_i)-V(\varphi_i^{\rm T})\right]} \,.
\end{align}
In order to compare $2\sigma_{\text{B}}$ with $\sigma_{\text{DW}}$, we add another path connecting ${\rm T}$ and ${\rm F_2}$. Therefore,
\begin{align}
    2\,\sigma_{\rm B}=&\int^{\varphi_1^{\rm F_1}}_{0}\dd\varphi_1~\sqrt{\sum_{i=1}^n\left(\frac{\dd\varphi_i^{\rm F_1}}{\dd\varphi_1^{\rm F_1}}\right)^2}\sqrt{2[V(\varphi_i^{\rm F_1})-V(\phi_i^{\rm T})]}\nonumber\\
    &+\int^{\varphi_1^{\rm F_2}}_{0}(-\dd\varphi_1)~ \sqrt{\sum_{i=1}^n\left(\frac{\dd\varphi_i^{\rm F_2}}{\dd\varphi_1^{\rm F_2}}\right)^2}\sqrt{2[V(\varphi_i^{\rm F_2})-V(\varphi_i^{\rm T})]}\,,
\end{align}
where the minus sign in second line accounts for $\varphi_1=a<0$ along the path. If we define the path connecting ${\rm F_1}$ and ${\rm F_2}$ by $\varphi_i^{\rm B}=\begin{cases}
    \varphi_i^{\rm F_2}~~\varphi_1\in[\varphi_1^{\rm F_2},0]\\
    \varphi_i^{\rm F_1}~~\varphi_1\in[0,\varphi_1^{\rm F_1}]
\end{cases}$, which corresponds to following the two sides of the triangle between ${\rm F_1}$ and ${\rm F_2}$ rather than the straight line connecting them. The tension is
\begin{align}
    2\sigma_{\rm B}=\int^{\varphi_1^{\rm F_1}}_{\varphi_1^{\rm F_2}}\dd\varphi_1\sqrt{\sum_{i=1}^n\left(\frac{\dd\varphi_i^{\rm B}}{\dd\varphi_1^{\rm B}}\right)^2}\sqrt{2[V(\varphi_i^{\rm B})-V(\varphi_i^{\rm T})]}\,.\label{eqn:2sigmaB_general}
\end{align}
Note that the domain-wall tension $\sigma_{\rm DW}$ can be written in exactly the same form as the right side of Eq.~\eqref{eqn:2sigmaB_general}, except that the fields follow the domain-wall trajectory $\varphi_i^{\rm DW}$. As long as the domain wall is stable, as shown in Appendix~\ref{app:differentZ2},the path $\varphi_i^{\rm DW}$ corresponds to a local minimum of the one-dimensional action and therefore minimizes the tension. Consequently, one obtains $\sigma_{\text{DW}} < 2\sigma_{\text{B}}$. However, this does not rule out complete wetting, because the ratio $\sigma_{\text{DW}} / 2\sigma_{\text{B}}$ appears only within the thin-wall approximation, and in some theories—such as the $SU(3)_c$ Polyakov-loop potential—the ${\rm T-F}$ and ${\rm F-F}$ wall thicknesses can be comparable to the size of the true vacuum, as shown in Refs.~\cite{Frei:1989es, Trappenberg:1992bt, Holland:2000uj}.


\begin{thebibliography}{10}

\bibitem{Kuzmin:1985mm}
V.~A. Kuzmin, V.~A. Rubakov and M.~E. Shaposhnikov, \emph{{On the Anomalous
  Electroweak Baryon Number Nonconservation in the Early Universe}},
  \href{https://doi.org/10.1016/0370-2693(85)91028-7}{\emph{Phys. Lett. B}
  {\bfseries 155} (1985) 36}.

\bibitem{Cohen:1993nk}
A.~G. Cohen, D.~B. Kaplan and A.~E. Nelson, \emph{{Progress in electroweak
  baryogenesis}},
  \href{https://doi.org/10.1146/annurev.ns.43.120193.000331}{\emph{Ann. Rev.
  Nucl. Part. Sci.} {\bfseries 43} (1993) 27--70},
  [\href{https://arxiv.org/abs/hep-ph/9302210}{{\ttfamily hep-ph/9302210}}].

\bibitem{Rubakov:1996vz}
V.~A. Rubakov and M.~E. Shaposhnikov, \emph{{Electroweak baryon number
  nonconservation in the early universe and in high-energy collisions}},
  \href{https://doi.org/10.1070/PU1996v039n05ABEH000145}{\emph{Usp. Fiz. Nauk}
  {\bfseries 166} (1996) 493--537},
  [\href{https://arxiv.org/abs/hep-ph/9603208}{{\ttfamily hep-ph/9603208}}].

\bibitem{LIGOScientific:2016aoc}
{\scshape LIGO Scientific, Virgo} collaboration, B.~P. Abbott et~al.,
  \emph{{Observation of Gravitational Waves from a Binary Black Hole Merger}},
  \href{https://doi.org/10.1103/PhysRevLett.116.061102}{\emph{Phys. Rev. Lett.}
  {\bfseries 116} (2016) 061102},
  [\href{https://arxiv.org/abs/1602.03837}{{\ttfamily 1602.03837}}].

\bibitem{ET:2019dnz}
{\scshape ET} collaboration, M.~Maggiore et~al., \emph{{Science Case for the
  Einstein Telescope}},
  \href{https://doi.org/10.1088/1475-7516/2020/03/050}{\emph{JCAP} {\bfseries
  03} (2020) 050}, [\href{https://arxiv.org/abs/1912.02622}{{\ttfamily
  1912.02622}}].

\bibitem{Caprini:2019egz}
C.~Caprini et~al., \emph{{Detecting gravitational waves from cosmological phase
  transitions with LISA: an update}},
  \href{https://doi.org/10.1088/1475-7516/2020/03/024}{\emph{JCAP} {\bfseries
  03} (2020) 024}, [\href{https://arxiv.org/abs/1910.13125}{{\ttfamily
  1910.13125}}].

\bibitem{TianQin:2015yph}
{\scshape TianQin} collaboration, J.~Luo et~al., \emph{{TianQin: a space-borne
  gravitational wave detector}},
  \href{https://doi.org/10.1088/0264-9381/33/3/035010}{\emph{Class. Quant.
  Grav.} {\bfseries 33} (2016) 035010},
  [\href{https://arxiv.org/abs/1512.02076}{{\ttfamily 1512.02076}}].

\bibitem{Hu:2017mde}
W.-R. Hu and Y.-L. Wu, \emph{{The Taiji Program in Space for gravitational wave
  physics and the nature of gravity}},
  \href{https://doi.org/10.1093/nsr/nwx116}{\emph{Natl. Sci. Rev.} {\bfseries
  4} (2017) 685--686}.

\bibitem{NANOGrav:2023gor}
{\scshape NANOGrav} collaboration, G.~Agazie et~al., \emph{{The NANOGrav 15 yr
  Data Set: Evidence for a Gravitational-wave Background}},
  \href{https://doi.org/10.3847/2041-8213/acdac6}{\emph{Astrophys. J. Lett.}
  {\bfseries 951} (2023) L8},
  [\href{https://arxiv.org/abs/2306.16213}{{\ttfamily 2306.16213}}].

\bibitem{Athron:2023xlk}
P.~Athron, C.~Bal{\'a}zs, A.~Fowlie, L.~Morris and L.~Wu, \emph{{Cosmological
  phase transitions: From perturbative particle physics to gravitational
  waves}}, \href{https://doi.org/10.1016/j.ppnp.2023.104094}{\emph{Prog. Part.
  Nucl. Phys.} {\bfseries 135} (2024) 104094},
  [\href{https://arxiv.org/abs/2305.02357}{{\ttfamily 2305.02357}}].

\bibitem{Coleman:1977py}
S.~R. Coleman, \emph{{The Fate of the False Vacuum. 1. Semiclassical Theory}},
  \href{https://doi.org/10.1103/PhysRevD.16.1248}{\emph{Phys. Rev. D}
  {\bfseries 15} (1977) 2929--2936}.

\bibitem{Callan:1977pt}
C.~G. Callan, Jr. and S.~R. Coleman, \emph{{The Fate of the False Vacuum. 2.
  First Quantum Corrections}},
  \href{https://doi.org/10.1103/PhysRevD.16.1762}{\emph{Phys. Rev. D}
  {\bfseries 16} (1977) 1762--1768}.

\bibitem{Linde:1977mm}
A.~D. Linde, \emph{{On the Vacuum Instability and the Higgs Meson Mass}},
  \href{https://doi.org/10.1016/0370-2693(77)90664-5}{\emph{Phys. Lett. B}
  {\bfseries 70} (1977) 306--308}.

\bibitem{Linde:1980tt}
A.~D. Linde, \emph{{Fate of the False Vacuum at Finite Temperature: Theory and
  Applications}},
  \href{https://doi.org/10.1016/0370-2693(81)90281-1}{\emph{Phys. Lett. B}
  {\bfseries 100} (1981) 37--40}.

\bibitem{Linde:1981zj}
A.~D. Linde, \emph{{Decay of the False Vacuum at Finite Temperature}},
  \href{https://doi.org/10.1016/0550-3213(83)90072-X}{\emph{Nucl. Phys. B}
  {\bfseries 216} (1983) 421}.

\bibitem{Guth:1981uk}
A.~H. Guth and E.~J. Weinberg, \emph{{Cosmological Consequences of a First
  Order Phase Transition in the SU(5) Grand Unified Model}},
  \href{https://doi.org/10.1103/PhysRevD.23.876}{\emph{Phys. Rev. D} {\bfseries
  23} (1981) 876}.

\bibitem{Volmer1929BERKU}
M.~Volmer, \emph{{\"U}ber keimbildung und keimwirkung als spezialf{\"a}lle der
  heterogenen katalyse}, {\emph{Zeitschrift f{\"u}r Elektrochemie und
  angewandte physikalische Chemie} (1929) }.

\bibitem{Volmer_book}
M.~Volmer, \emph{Kinetik der Phasenbildung (Kinetics of Phase Formation)}.
\newblock Dresden and Leipzig, 1939, available at:
  \url{https://apps.dtic.mil/sti/citations/tr/ADA800534}.

\bibitem{Turnbull_Fisher}
D.~Turnbull and J.~C. Fisher, \emph{Rate of nucleation in condensed systems},
  \href{https://doi.org/10.1063/1.1747055}{\emph{The Journal of Chemical
  Physics} {\bfseries 17} (01, 1949) 71--73},
  [\href{https://arxiv.org/abs/https://pubs.aip.org/aip/jcp/article-pdf/17/1/71/18795584/71\_1\_online.pdf}{{\ttfamily
  https://pubs.aip.org/aip/jcp/article-pdf/17/1/71/18795584/71\_1\_online.pdf}}].

\bibitem{Turnbull}
D.~Turnbull, \emph{Formation of crystal nuclei in liquid metals},
  \href{https://doi.org/10.1063/1.1699435}{\emph{Journal of Applied Physics}
  {\bfseries 21} (10, 1950) 1022--1028},
  [\href{https://arxiv.org/abs/https://pubs.aip.org/aip/jap/article-pdf/21/10/1022/18309730/1022\_1\_online.pdf}{{\ttfamily
  https://pubs.aip.org/aip/jap/article-pdf/21/10/1022/18309730/1022\_1\_online.pdf}}].

\bibitem{mullin2001crystallization}
J.~Mullin, \emph{Crystallization}.
\newblock Chemical, Petrochemical \& Process. Butterworth-Heinemann, 2001.

\bibitem{Steinhardt:1981ec}
P.~J. Steinhardt, \emph{{Monopole and Vortex Dissociation and Decay of the
  False Vacuum}},
  \href{https://doi.org/10.1016/0550-3213(81)90449-1}{\emph{Nucl. Phys. B}
  {\bfseries 190} (1981) 583--616}.

\bibitem{Hosotani:1982ii}
Y.~Hosotani, \emph{{Impurities in the Early Universe}},
  \href{https://doi.org/10.1103/PhysRevD.27.789}{\emph{Phys. Rev. D} {\bfseries
  27} (1983) 789}.

\bibitem{Yajnik:1986tg}
U.~A. Yajnik, \emph{{PHASE TRANSITION INDUCED BY COSMIC STRINGS}},
  \href{https://doi.org/10.1103/PhysRevD.34.1237}{\emph{Phys. Rev. D}
  {\bfseries 34} (1986) 1237--1240}.

\bibitem{Frei:1989es}
Z.~Frei and A.~Patkos, \emph{{Perfect Wetting: An Alternative for Hadronic
  Matter Formation in the Cooling Universe}},
  \href{https://doi.org/10.1016/0370-2693(89)90164-0}{\emph{Phys. Lett. B}
  {\bfseries 229} (1989) 102--106}.

\bibitem{Trappenberg:1992bt}
T.~Trappenberg and U.~J. Wiese, \emph{{Z(3) instantons in models for wetting of
  hot gluons}}, \href{https://doi.org/10.1016/0550-3213(92)90372-I}{\emph{Nucl.
  Phys. B} {\bfseries 372} (1992) 703--726}.

\bibitem{Christiansen:1995ic}
M.~B. Christiansen and J.~Madsen, \emph{{Large nucleation distances from
  impurities in the cosmological quark - hadron transition}},
  \href{https://doi.org/10.1103/PhysRevD.53.5446}{\emph{Phys. Rev. D}
  {\bfseries 53} (1996) 5446--5454},
  [\href{https://arxiv.org/abs/astro-ph/9602071}{{\ttfamily
  astro-ph/9602071}}].

\bibitem{Holland:2000uj}
K.~Holland and U.-J. Wiese, \emph{{The Center symmetry and its spontaneous
  breakdown at high temperatures}},
  \href{https://arxiv.org/abs/hep-ph/0011193}{{\ttfamily hep-ph/0011193}}.

\bibitem{Hiscock:1987hn}
W.~A. Hiscock, \emph{{CAN BLACK HOLES NUCLEATE VACUUM PHASE TRANSITIONS?}},
  \href{https://doi.org/10.1103/PhysRevD.35.1161}{\emph{Phys. Rev. D}
  {\bfseries 35} (1987) 1161--1170}.

\bibitem{Burda:2015isa}
P.~Burda, R.~Gregory and I.~Moss, \emph{{Gravity and the stability of the Higgs
  vacuum}}, \href{https://doi.org/10.1103/PhysRevLett.115.071303}{\emph{Phys.
  Rev. Lett.} {\bfseries 115} (2015) 071303},
  [\href{https://arxiv.org/abs/1501.04937}{{\ttfamily 1501.04937}}].

\bibitem{Oshita:2018ptr}
N.~Oshita, M.~Yamada and M.~Yamaguchi, \emph{{Compact objects as the catalysts
  for vacuum decays}},
  \href{https://doi.org/10.1016/j.physletb.2019.02.032}{\emph{Phys. Lett. B}
  {\bfseries 791} (2019) 149--155},
  [\href{https://arxiv.org/abs/1808.01382}{{\ttfamily 1808.01382}}].

\bibitem{Grinstein:2015jda}
B.~Grinstein and C.~W. Murphy, \emph{{Semiclassical Approach to Heterogeneous
  Vacuum Decay}}, \href{https://doi.org/10.1007/JHEP12(2015)063}{\emph{JHEP}
  {\bfseries 12} (2015) 063},
  [\href{https://arxiv.org/abs/1509.05405}{{\ttfamily 1509.05405}}].

\bibitem{Blasi:2022woz}
S.~Blasi and A.~Mariotti, \emph{{Domain Walls Seeding the Electroweak Phase
  Transition}},
  \href{https://doi.org/10.1103/PhysRevLett.129.261303}{\emph{Phys. Rev. Lett.}
  {\bfseries 129} (2022) 261303},
  [\href{https://arxiv.org/abs/2203.16450}{{\ttfamily 2203.16450}}].

\bibitem{Agrawal:2023cgp}
P.~Agrawal, S.~Blasi, A.~Mariotti and M.~Nee, \emph{{Electroweak phase
  transition with a double well done doubly well}},
  \href{https://doi.org/10.1007/JHEP06(2024)089}{\emph{JHEP} {\bfseries 06}
  (2024) 089}, [\href{https://arxiv.org/abs/2312.06749}{{\ttfamily
  2312.06749}}].

\bibitem{Blasi:2024mtc}
S.~Blasi and A.~Mariotti, \emph{{QCD axion strings or seeds?}},
  \href{https://doi.org/10.21468/SciPostPhys.18.1.016}{\emph{SciPost Phys.}
  {\bfseries 18} (2025) 016},
  [\href{https://arxiv.org/abs/2405.08060}{{\ttfamily 2405.08060}}].

\bibitem{PinaAvelino:2006ia}
P.~Pina~Avelino, C.~J. A.~P. Martins, J.~Menezes, R.~Menezes and J.~C. R.~E.
  Oliveira, \emph{{Frustrated expectations: defect networks and dark energy}},
  \href{https://doi.org/10.1103/PhysRevD.73.123519}{\emph{Phys. Rev. D}
  {\bfseries 73} (2006) 123519},
  [\href{https://arxiv.org/abs/astro-ph/0602540}{{\ttfamily
  astro-ph/0602540}}].

\bibitem{Avelino:2006xf}
P.~P. Avelino, C.~J. A.~P. Martins, J.~Menezes, R.~Menezes and J.~C. R.~E.
  Oliveira, \emph{{Scaling of cosmological domain wall networks with
  junctions}},
  \href{https://doi.org/10.1016/j.physletb.2007.02.025}{\emph{Phys. Lett. B}
  {\bfseries 647} (2007) 63--66},
  [\href{https://arxiv.org/abs/astro-ph/0612444}{{\ttfamily
  astro-ph/0612444}}].

\bibitem{Avelino:2008ve}
P.~P. Avelino, C.~J. A.~P. Martins, J.~Menezes, R.~Menezes and J.~C. R.~E.
  Oliveira, \emph{{Dynamics of domain wall networks with junctions}},
  \href{https://doi.org/10.1103/PhysRevD.78.103508}{\emph{Phys. Rev. D}
  {\bfseries 78} (2008) 103508},
  [\href{https://arxiv.org/abs/0807.4442}{{\ttfamily 0807.4442}}].

\bibitem{Battye:2011ff}
R.~A. Battye, J.~A. Pearson and A.~Moss, \emph{{X-type and Y-type junction
  stability in domain wall networks}},
  \href{https://doi.org/10.1103/PhysRevD.84.125032}{\emph{Phys. Rev. D}
  {\bfseries 84} (2011) 125032},
  [\href{https://arxiv.org/abs/1107.1325}{{\ttfamily 1107.1325}}].

\bibitem{Abraham:1990nz}
E.~R.~C. Abraham and P.~K. Townsend, \emph{{Intersecting extended objects in
  supersymmetric field theories}},
  \href{https://doi.org/10.1016/0550-3213(91)90093-D}{\emph{Nucl. Phys. B}
  {\bfseries 351} (1991) 313--332}.

\bibitem{Gibbons:1999np}
G.~W. Gibbons and P.~K. Townsend, \emph{{A Bogomolny equation for intersecting
  domain walls}},
  \href{https://doi.org/10.1103/PhysRevLett.83.1727}{\emph{Phys. Rev. Lett.}
  {\bfseries 83} (1999) 1727--1730},
  [\href{https://arxiv.org/abs/hep-th/9905196}{{\ttfamily hep-th/9905196}}].

\bibitem{Carroll:1999wr}
S.~M. Carroll, S.~Hellerman and M.~Trodden, \emph{{Domain wall junctions are
  1/4 - BPS states}},
  \href{https://doi.org/10.1103/PhysRevD.61.065001}{\emph{Phys. Rev. D}
  {\bfseries 61} (2000) 065001},
  [\href{https://arxiv.org/abs/hep-th/9905217}{{\ttfamily hep-th/9905217}}].

\bibitem{Agrawal:2022hnf}
P.~Agrawal and M.~Nee, \emph{{The Boring Monopole}},
  \href{https://doi.org/10.21468/SciPostPhys.13.3.049}{\emph{SciPost Phys.}
  {\bfseries 13} (2022) 049},
  [\href{https://arxiv.org/abs/2202.11102}{{\ttfamily 2202.11102}}].

\bibitem{young1805}
T.~Young, \emph{An essay on the cohesion of fluids}, {\emph{Philosophical
  Transactions of the Royal Society of London} {\bfseries 95} (1805) 65--87}.

\bibitem{gibbs1957collected}
J.~W. J.~W. Gibbs, \emph{The Collected Works of J. Willard Gibbs}.
\newblock Yale University Press, New Haven, London, 1957.

\bibitem{Ellis:2018mja}
J.~Ellis, M.~Lewicki and J.~M. No, \emph{{On the Maximal Strength of a
  First-Order Electroweak Phase Transition and its Gravitational Wave Signal}},
  \href{https://doi.org/10.1088/1475-7516/2019/04/003}{\emph{JCAP} {\bfseries
  04} (2019) 003}, [\href{https://arxiv.org/abs/1809.08242}{{\ttfamily
  1809.08242}}].

\bibitem{erdos59a}
P.~Erd\"{o}s and A.~R\'{e}nyi, \emph{On random graphs i}, {\emph{Publicationes
  Mathematicae Debrecen} {\bfseries 6} (1959) 290}.

\bibitem{PhysRevB.30.3933}
I.~Balberg, C.~H. Anderson, S.~Alexander and N.~Wagner, \emph{Excluded volume
  and its relation to the onset of percolation},
  \href{https://doi.org/10.1103/PhysRevB.30.3933}{\emph{Phys. Rev. B}
  {\bfseries 30} (Oct, 1984) 3933--3943}.

\bibitem{PhysRevE.76.051115}
J.~A. Quintanilla and R.~M. Ziff, \emph{Asymmetry in the percolation thresholds
  of fully penetrable disks with two different radii},
  \href{https://doi.org/10.1103/PhysRevE.76.051115}{\emph{Phys. Rev. E}
  {\bfseries 76} (Nov, 2007) 051115}.

\bibitem{Bai:2022kxq}
Y.~Bai, S.~Lu and N.~Orlofsky, \emph{{Origin of nontopological soliton dark
  matter: solitosynthesis or phase transition}},
  \href{https://doi.org/10.1007/JHEP10(2022)181}{\emph{JHEP} {\bfseries 10}
  (2022) 181}, [\href{https://arxiv.org/abs/2208.12290}{{\ttfamily
  2208.12290}}].

\bibitem{Kibble:1976sj}
T.~W.~B. Kibble, \emph{Topology of cosmic domains and strings},
  \href{https://doi.org/10.1088/0305-4470/9/8/029}{\emph{J. Phys. A} {\bfseries
  9} (1976) 1387}.

\bibitem{Zurek:1985qw}
W.~H. Zurek, \emph{Cosmological experiments in superfluid helium?},
  \href{https://doi.org/10.1038/317505a0}{\emph{Nature} {\bfseries 317} (1985)
  505}.

\bibitem{Leite:2011np}
A.~M.~M. Leite and C.~J. A.~P. Martins, \emph{Scaling properties of domain wall
  networks}, \href{https://doi.org/10.1103/PhysRevD.84.103523}{\emph{Phys. Rev.
  D} {\bfseries 84} (2011) 103523},
  [\href{https://arxiv.org/abs/1110.3486}{{\ttfamily 1110.3486}}].

\bibitem{Guada:2020xnz}
V.~Guada, M.~Nemev{\v{s}}ek and M.~Pintar, \emph{{FindBounce: Package for
  multi-field bounce actions}},
  \href{https://doi.org/10.1016/j.cpc.2020.107480}{\emph{Comput. Phys. Commun.}
  {\bfseries 256} (2020) 107480},
  [\href{https://arxiv.org/abs/2002.00881}{{\ttfamily 2002.00881}}].

\bibitem{Wainwright:2011kj}
C.~L. Wainwright, \emph{{CosmoTransitions: Computing Cosmological Phase
  Transition Temperatures and Bubble Profiles with Multiple Fields}},
  \href{https://doi.org/10.1016/j.cpc.2012.04.004}{\emph{Comput. Phys. Commun.}
  {\bfseries 183} (2012) 2006--2013},
  [\href{https://arxiv.org/abs/1109.4189}{{\ttfamily 1109.4189}}].

\bibitem{AMBROSETTI1973349}
A.~Ambrosetti and P.~H. Rabinowitz, \emph{Dual variational methods in critical
  point theory and applications},
  \href{https://doi.org/https://doi.org/10.1016/0022-1236(73)90051-7}{\emph{Journal
  of Functional Analysis} {\bfseries 14} (1973) 349--381}.

\bibitem{Wu:2022stu}
Y.~Wu, K.-P. Xie and Y.-L. Zhou, \emph{{Collapsing domain walls beyond Z2}},
  \href{https://doi.org/10.1103/PhysRevD.105.095013}{\emph{Phys. Rev. D}
  {\bfseries 105} (2022) 095013},
  [\href{https://arxiv.org/abs/2204.04374}{{\ttfamily 2204.04374}}].

\bibitem{Wu:2022tpe}
Y.~Wu, K.-P. Xie and Y.-L. Zhou, \emph{{Classification of Abelian domain
  walls}}, \href{https://doi.org/10.1103/PhysRevD.106.075019}{\emph{Phys. Rev.
  D} {\bfseries 106} (2022) 075019},
  [\href{https://arxiv.org/abs/2205.11529}{{\ttfamily 2205.11529}}].

\bibitem{Saffin:1999au}
P.~M. Saffin, \emph{{Tiling with almost BPS junctions}},
  \href{https://doi.org/10.1103/PhysRevLett.83.4249}{\emph{Phys. Rev. Lett.}
  {\bfseries 83} (1999) 4249--4252},
  [\href{https://arxiv.org/abs/hep-th/9907066}{{\ttfamily hep-th/9907066}}].

\bibitem{Blasi:2023rqi}
S.~Blasi, R.~Jinno, T.~Konstandin, H.~Rubira and I.~Stomberg,
  \emph{{Gravitational waves from defect-driven phase transitions: domain
  walls}}, \href{https://doi.org/10.1088/1475-7516/2023/10/051}{\emph{JCAP}
  {\bfseries 10} (2023) 051},
  [\href{https://arxiv.org/abs/2302.06952}{{\ttfamily 2302.06952}}].

\bibitem{Azzola:2024pzq}
J.~Azzola, O.~Matsedonskyi and A.~Weiler, \emph{{Minimal electroweak
  baryogenesis via domain walls}},
  \href{https://doi.org/10.1007/JHEP04(2025)103}{\emph{JHEP} {\bfseries 04}
  (2025) 103}, [\href{https://arxiv.org/abs/2412.10495}{{\ttfamily
  2412.10495}}].

\bibitem{Battye:2025uny}
R.~A. Battye, S.~J. Cotterill and A.~K. Thomasson, \emph{{Complete
  Classification of Domain Wall Solutions in the $\mathbb{Z}_2$-symmetric
  2HDM}},  \href{https://arxiv.org/abs/2509.23332}{{\ttfamily 2509.23332}}.

\end{thebibliography}

\providecommand{\href}[2]{#2}\begingroup\raggedright\endgroup

\end{document}